\documentclass[
reprint,
superscriptaddress,
amsmath,amssymb,
aps,
pra,
floatfix,
]{revtex4-2}
\pdfoutput=1
\usepackage{xcolor}
\usepackage{graphicx}
\usepackage{dcolumn}
\usepackage{bm}
\usepackage{filecontents}
\begin{document}

\title{Memory effects on the current induced propagation of spin textures in NdCo$_5$/Ni$_8$Fe$_2$ bilayers}

\author{V.V. Fernández}
\affiliation{Depto. Física, Universidad de Oviedo, 33007 Oviedo, Spain.\\} 
\affiliation{CINN (CSIC–Universidad de Oviedo), El Entrego, Spain.\\}
\author{A.E. Herguedas-Alonso}
\affiliation{Depto. Física, Universidad de Oviedo, 33007 Oviedo, Spain.\\} 
\affiliation{CINN (CSIC–Universidad de Oviedo), El Entrego, Spain.\\}
\affiliation{ALBA Synchrotron, 08290 Cerdanyola del Vallès, Spain.\\}
\author{J. Hermosa}
\affiliation{Depto. Física, Universidad de Oviedo, 33007 Oviedo, Spain.\\} 
\affiliation{CINN (CSIC–Universidad de Oviedo), El Entrego, Spain.\\}
\author{L.Aballe}
\affiliation{ALBA Synchrotron, 08290 Cerdanyola del Vallès, Spain.\\}
\author{A. Sorrentino}
\affiliation{ALBA Synchrotron, 08290 Cerdanyola del Vallès, Spain.\\}
\author{R. Valcarcel}
\affiliation{ALBA Synchrotron, 08290 Cerdanyola del Vallès, Spain.\\}
\author{C. Quiros}
\affiliation{Depto. Física, Universidad de Oviedo, 33007 Oviedo, Spain.\\} 
\affiliation{CINN (CSIC–Universidad de Oviedo), El Entrego, Spain.\\}
\author{J. I. Martín}
\affiliation{Depto. Física, Universidad de Oviedo, 33007 Oviedo, Spain.\\} 
\affiliation{CINN (CSIC–Universidad de Oviedo), El Entrego, Spain.\\}
\author{E. Pereiro}
\affiliation{ALBA Synchrotron, 08290 Cerdanyola del Vallès, Spain.\\}
\author{S. Ferrer}
\affiliation{ALBA Synchrotron, 08290 Cerdanyola del Vallès, Spain.\\}
\author{A. Hierro-Rodríguez}
\affiliation{Depto. Física, Universidad de Oviedo, 33007 Oviedo, Spain.\\} 
\affiliation{CINN (CSIC–Universidad de Oviedo), El Entrego, Spain.\\}
\affiliation{SUPA, School of Physics and Astronomy, University of Glasgow G12 8QQ, UK.\\}
\author{M. Vélez}
\affiliation{Depto. Física, Universidad de Oviedo, 33007 Oviedo, Spain.\\} 
\affiliation{CINN (CSIC–Universidad de Oviedo), El Entrego, Spain.\\}

\date{\today}

\begin{abstract}

Bilayers of NdCo$_5$/Ni$_8$Fe$_2$ can act as reconfigurable racetracks thanks to the parallel stripe domain configuration present in the hard magnetic material with weak perpendicular anisotropy (NdCo$_5$), and its imprint on the soft magnetic layer (Ni$_8$Fe$_2$). This pattern hosts spin textures with well defined topological charges and establishes paths for their deterministic propagation under the effect of pulsed currents, which has been studied as a function of externally applied fields by using Magnetic Transmission X-ray Microscopy. The experiments show guided vortex/antivortex propagation events within the Ni$_8$Fe$_2$ above a threshold current of $3\cdot10^{11}$ A/m$^2$. Opposite propagation senses have been observed depending on the topological charge of the spin texture, both in the remnant state and under an applied external field. Micromagnetic simulations of our multilayer reveal that the guiding effect and asymmetric propagation sense are due to the magnetic history of the hard magnetic layer. An exchange-bias-based memory effect acts as a magnetic spring and controls the propagation sense by favoring a specific orientation of the in plane magnetization, leading to a system which behaves as a hard-soft magnetic composite with reconfigurable capabilities for a controlled propagation of magnetic topological textures. 
\end{abstract}

\maketitle

\section{\label{sec:Intro} Introduction}

Domain wall (DW) racetracks are basic elements in the design of energy efficient spintronic devices for applications ranging from high density   recording \cite{Parkin} to stochastic \cite{Dedalo} and quantum \cite{quantum} computing architectures. A key issue in the magnetic racetrack concept is the current induced propagation of spin textures in restricted geometries \cite{planar, DWracetrack, skyrmion, Schafer} with different effects at play such as spin transfer \cite{STT1, STT} and spin orbit torques \cite{SOT}, oersted fields \cite{Fruchart, nw, Lucas}, Joule heating \cite{Joule}, etc. 

Most studies rely on hard sample boundaries to guide the propagating spin textures (e.g. nanowires \cite{Fruchart, nw}, 3D structures \cite{Parkin, ACSNano22}, patterned wires \cite{Dedalo, skyrmion}, curved strips \cite{Schafer}) and focus on the motion of individual magnetic textures such as domain walls or skyrmions \cite{Parkin, DWracetrack, skyrmion}. However, in certain configurations, composite textures provide optimized propagation properties such as skyrmion pairs in synthetic antiferromagnets with reduced skyrmion Hall effect \cite{SAF} or bimerons in films with in-plane magnetization with enhanced propagation distances \cite{bimeron, bimeron2}. 

Soft patterning by dipolar and exchange interactions has been proposed as an alternative route to design reconfigurable magnetic devices in which texture nucleation \cite{dipolar} and/or local pinning potentials \cite{reconfig1, reconfig2} can be externally controlled and used to read previously recorded non-volatile magnetic states. In this context, guiding potentials of magnetic origin \cite{Italiano} could provide additional flexibility to define variable propagation paths for adjustable interconnects \cite{DWlogic}. For example, a reconfigurable racetrack could be designed by the combination of a hard magnetic layer (HML) with a domain configuration that imprints predefined paths for the propagation of spin textures on a neighboring soft magnetic layer (SML), adjustable by the magnetic history of the HML.

Weak Perpendicular Magnetic Anisotropy (PMA) Nd$_x$Co$_{1-x}$ alloys can be an interesting choice for the HML. These materials display a parallel stripe domain configuration \cite{huber} with a characteristic property of rotatable anisotropy \cite{PRB2013, OriginRotAnis}: stripe orientation can be externally adjusted by the direction of the last in-plane saturating field. This effect has been used to create confined labyrinths in patterned NdCo films \cite{PRL2012}, to induce exchange bias like effects in hard/soft lateral composites \cite{APL2014} and, more important, to imprint parallel domain configurations  in different in-plane magnetic systems as GdCo and NiFe alloys, by exchange and magnetostatic interactions \cite{NC2015, PRB2017}. 

Stripe patterns of weak PMA materials host spin textures with well defined topological charges due to the constraints imposed by the periodic domain configuration \cite{topological, NC2015}. In particular, the onset of in plane magnetization reversal takes place by the nucleation of magnetic bimerons at the sample surfaces \cite{10.1063/1.4984898, PRB2017}, which are imprinted into the SML of a HML/SML system. Each bimeron is a composite texture made of a vortex/antivortex (V/AV) pair with opposite out-of-plane polarity that carries a topological charge $Q$ equivalent to the skyrmion ($Q=\pm 1$) \cite{bimeron}. Pulsed field experiments on GdCo/NdCo/NiFe multilayers \cite{10.1063/1.4984898}, have reported the guided propagation of bimerons, individual Vs and individual AVs along the direction defined by the stripe domain pattern within a limited field range around coercivity \cite{10261_173549}. 

Here, the propagation of spin textures along the stripe domain pattern of a HML/SML bilayer (HML=NdCo$_5$, SML=NiFe) has been studied under the effect of pulsed currents as a function of the HML magnetic history and in-plane applied fields. Element resolved Magnetic Transmission X-ray microscopy (MTXM) experiments show guided V/AV propagation events within the SML over a threshold current of $10^{11}$ A/m$^2$. The propagation sense is found to be determined by the topological charge of each texture, keeping memory of the last saturated state in the HML due to exchange interactions at the HML/SML interface.

\section{\label{sec:Exp} Experimental}

\begin{figure}[ht]
    \centering
    \includegraphics[width = 1\linewidth]{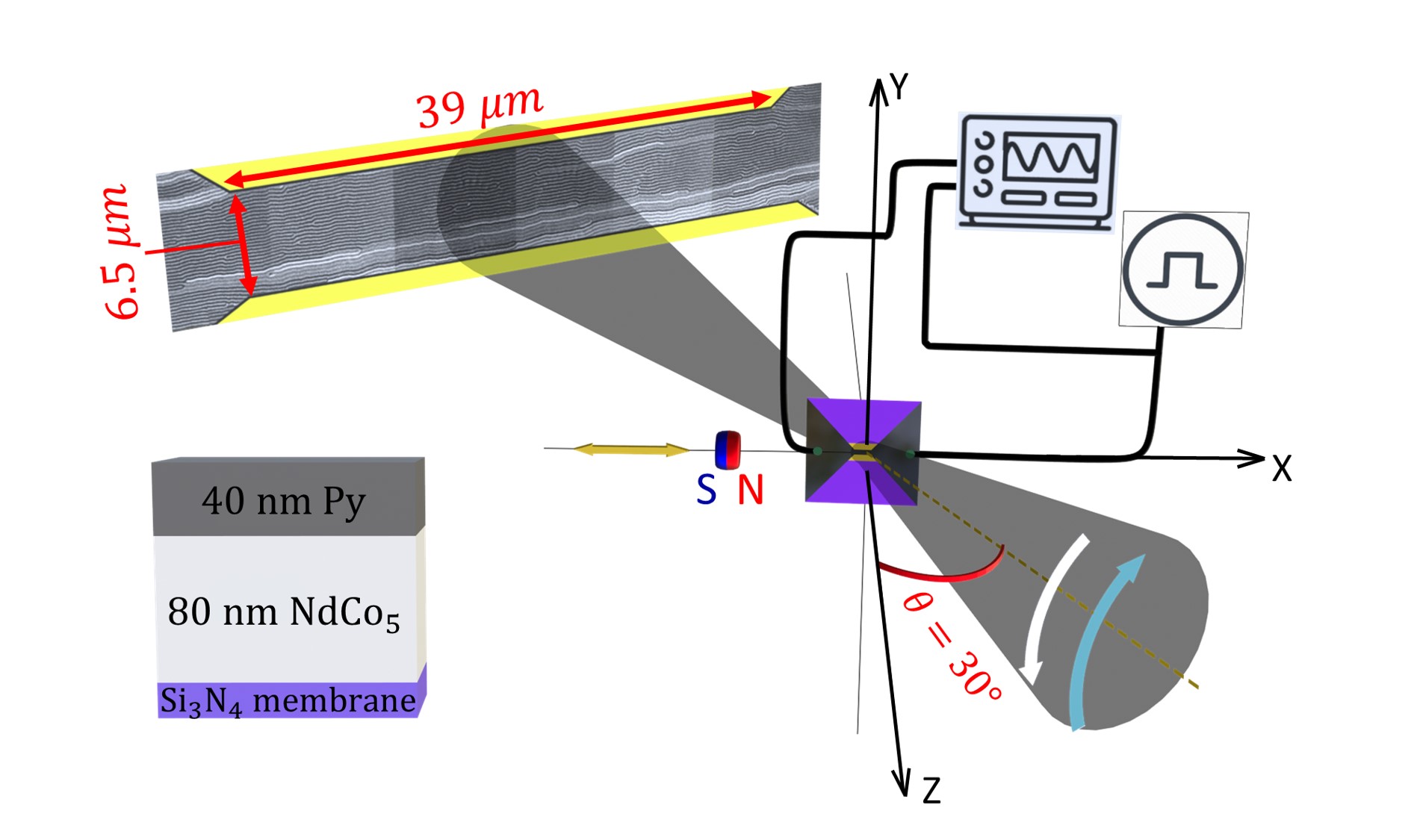}
        \caption{\label{fig:Figure1} Schematic representation of the experimental layout for MTXM measurements. A circularly polarized X-ray beam is used to image the magnetic configuration of a transport bridge connected to a pulse generator and an oscilloscope. The sample is rotated by $\theta$=30$^{\circ}$ around the Y-axis to attain sensitivity to both in-plane and out-of-plane magnetization components, and to be able to apply in-plane magnetic fields with the polar magnet. Constant fields parallel to the sample plane are applied with a permanent magnet, with variable intensity depending on the relative distance between magnet and sample (the double yellow arrow shows the direction in which the magnet moves). A typical MTXM image of the stripe domain configuration in the SML is shown at the detector plane. The inset is a sketch of the magnetic heterostructure.}
\end{figure}

The sample used in this study is a 80 nm NdCo$_5$/ \hspace{10mm}40 nm Ni$_8$Fe$_2$ HML/SML bilayer (as sketched in the inset of Fig. \ref{fig:Figure1}) fabricated by dc magnetron sputtering under similar conditions as reported previously \cite{10.1063/1.4984898}. The HML is made of NdCo$_5$, an amorphous alloy with moderate out of plane anisotropy $K_N\approx10^5$ J/m$^3$ \cite{10261_173549}. HML thickness is 80 nm, chosen to favor the generation of magnetic stripe domain patterns with out-of-plane up/down magnetization. The SML is made of  permalloy (Py = Ni$_8$Fe$_2$) with in-plane uniaxial anisotropy $K_u= 850$ J/m$^3$ and weak out of plane anisotropy $K_N\approx10^4$ J/m$^3$ \cite{10261_173549}. Our HML has high electrical resistivity \mbox{($\rho_{NdCo}=1.3\cdot10^{-6}$ $\Omega$m)} compared to that of the SML ($\rho_{Py}=0.3\cdot10^{-6}$ $\Omega$m). The large difference in resistivity between amorphous NdCo$_5$ and policrystalline Py ensures that transport currents will flow preferentially within the SML, that hosts the spin textures. The HML/SML bilayer was deposited onto a 200 nm thick Si$_3$N$_4$ X-ray transparent membrane on an intrinsic Si frame with a window size of 250 x 250 $\mu$m$^2$. A 39 $\mu$m long and 6.5 $\mu$m wide transport bridge was fabricated for the pulsed current experiments using optical lithography and lift off (Figure \ref{fig:Figure1}). A double layer was used by combining LOR-3B and S-1813 resists. This two layer procedure favors the generation of an undercut in the resist mask making the deposited thin film discontinuous so that the efficiency of the lift-off process on the Si$_3$N$_4$ membrane is improved.

The sample was mounted into the full field microscope of the MISTRAL beamline \cite{MISTRAL} at the ALBA Synchrotron for the MTXM experiments. Transmission X-ray microscopy images were acquired using right and left-handed circularly polarized light. Magnetic contrast images were obtained from the subtraction of the logarithm of individual transmittance images acquired with opposite X-ray helicity, exploiting the X-ray magnetic circular dichroism (XMCD) effect \cite{stöhr2023nature}. Photon energy was tuned at the Fe L$_3$ absorption edge (706.8 eV) in order to record the magnetic configuration of spin textures within the SML Ni$_8$Fe$_2$ alloy. The XMCD contrast is governed by the dot product of the spin angular momentum of the photons and the magnetization ($\boldsymbol{\sigma}\cdot\boldsymbol{m}$) \cite{stöhr2023nature}, i.e. image contrast is sensitive to the magnetization component parallel to the X-ray beam. For this reason the sample was rotated $\theta=30^{\circ}$ around the Y-axis, which is parallel to the sample surface and perpendicular to the horizontal X-ray beam (Figure \ref{fig:Figure1}), to be sensitive to both out of plane ($m_z$) and in plane ($m_x$) magnetization components (i.e. $\boldsymbol{\sigma}\cdot\boldsymbol{m} \propto m_x \sin(30^{\circ})+m_z \cos(30^{\circ})$). 

Two electrical contacts established on top of the Py layer allowed the injection of the electrical current. The sample was connected to a pulse generator (AVTECH AVR-E3-B-W3 pulse generator) and an oscilloscope to monitor the electrical signal. Then, in order to induce the controlled propagation of spin textures \cite{Fruchart}, short current pulses of different polarities ($\sim$ 20 ns wide) were applied to the transport bridge with variable amplitude up to a current density of 4.4$\cdot$10$^{11}$ A/m$^2$ within the SML. A permanent magnet with its field parallel to the sample's surface was used to apply static magnetic fields of variable intensity (from -50 mT to 50 mT), depending on the relative sample-magnet distance and magnet orientation (Figure \ref{fig:Figure1}). Consecutive MTXM images were acquired before and after each current pulse to assess the changes in the sample magnetic configuration.
\begin{figure*}[ht]
    \centering
    \includegraphics[width = 1\linewidth]{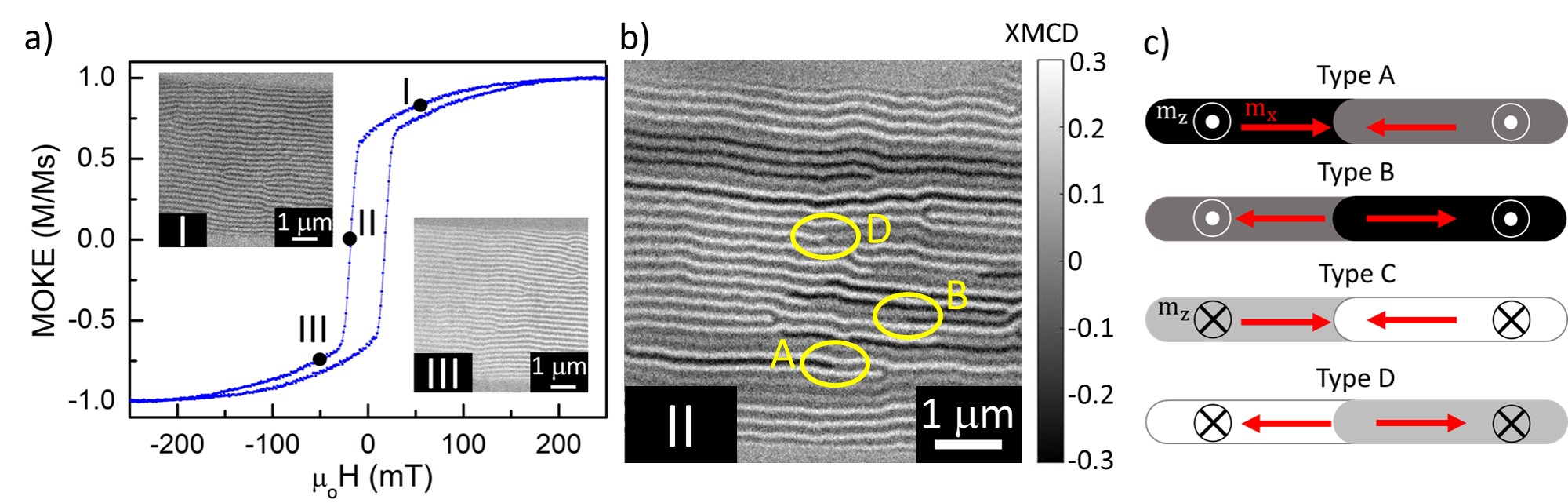}
    \caption{\label{fig:Figure2} Hysteresis loop and MTXM of 80 nm NdCo$_5$/40 nm Py bilayer. (a) In-plane TMOKE hysteresis loop. The insets show MTXM images of the SML at $\mu_0H = 50$ mT (state I) and -50 mT (state III). (b) MTXM image at the coercive field $\mu_0H \approx -18$ mT (state II). (c) Schematic representation of the relative orientations of $m_x$ (red arrows) and $m_z$ (black and white circles) and the corresponding contrast levels. Classification of the four different types of domain walls: Types A and C have a head-to-head $m_x$ orientation and correspond to spin textures with $Q=-1/2$, while types B and D have a tail-to-tail orientation corresponding to spin textures with $Q=+1/2$. Three examples of these structures are marked with yellow circles in panel (b). }
\end{figure*}

Micromagnetic simulations of the HML/SML bilayer were performed using Mumax$^3$ \cite{Mumax} with simulation size $3072\times2110\times120$ nm$^3$, a cell size of $4\times4.12\times4$ nm$^3$ and periodic boundary conditions along the $X$ direction. Material parameters were extracted from the magnetic characterization of control samples by Vibrating Sample Magnetometry (VSM) and Magnetooptical Transverse Kerr Effect (TMOKE). The 80 nm thick HML NdCo$_5$ alloy was simulated with a $M_S=8\cdot10^5$ A/m, $K_{OP}=1.35\cdot10^5$ J/m$^3$, $K_{IP}=4\cdot10^3$ J/m$^3$ and exchange stiffness $A=1.3\cdot10^{-11}$ J/m. The 40 nm thick SML layer has $M_S=8\cdot10^5$ A/m, $K_{OP}=1.1\cdot10^4$ J/m$^3$, $K_{IP}=850$ J/m$^3$ and exchange stiffness $A=1.3\cdot10^{-11}$ J/m. In a first step, we determined the equilibrium stripe period at remanence of the HML/SML bilayer from the relaxed state with minimum energy, starting from positive saturation along $X$ direction. In a second step, bifurcations were nucleated defining two or more different regions where different periodicities were imposed, in order to force the system to create the defects when relaxed at zero field. Then, a small negative external magnetic field around $\mu_0$H$_{applied}\approx$ 10 mT was applied, opposite to the initial saturating sense of the magnetization. This field is around half the coercive field and favors partial reversal of the magnetization, generating the desired bimerons in the Py layer. Finally, this field was removed and the system was allowed to relax resulting in a remnant state with partial inversion of the magnetization and V/AV pairs. 

\section{\label{sec:RyD} Results and discussion}

\subsection{\label{sec:MCSML} Magnetic configuration and spin textures within the SML}

The TMOKE hysteresis loop of the HML/SML bilayer (Figure $\ref{fig:Figure2}$(a)) shows the typical transcritical shape of stripe domain systems \cite{PRB2013} with a saturation field  of $\mu_0H_{sat}= 200$ mT, coercivity at $\mu_0H_c = 18$ mT and a reduced in-plane remanence, $M_r\approx 0.7 M_S$. MTXM images of the SML at positions I and III of the loop ($\pm 50$ mT, respectively) show a characteristic pattern of light/dark parallel bands with period $\approx 210$ nm which confirms that the stripe pattern of $\pm |m_z|$ domains has been imprinted  into the SML. There is, however, a clear difference between the magnetic contrast in states I and III: the stripes are black/light grey in the MTXM image at +50 mT and become dark gray/white at -50 mT. This is a signature of the different sign for the in-plane magnetization component \cite{NC2015}: at state I, the $\pm |m_z|$ oscillation occurs around a positive $m_x$ (resulting in a larger XMCD absorption  $\boldsymbol{\sigma}\cdot\boldsymbol{m} \propto +|m_x| \sin(30^{\circ})\pm |m_z| \cos(30^{\circ})$ whereas at state III, the average in-plane magnetization is negative (resulting in a smaller $\boldsymbol{\sigma}\cdot\boldsymbol{m} \propto -|m_x| \sin(30^{\circ})\pm |m_z| \cos(30^{\circ})$). 

At coercivity (state II with $\mu_0H_c=-18$ mT), four different intensity levels can be seen in the MTXM image (Fig. $\ref{fig:Figure2}$(b)) depending on the four possible sign combinations of $\pm |m_x|$ and $\pm |m_z|$ domains. Additionally, domain walls (DW) appear within individual stripe domains, characterized by a $m_x$ reversal in the middle of a stripe domain where $m_z$ remains constant (see yellow circles in Fig. $\ref{fig:Figure2}$(b)). We can distinguish four types of DWs depending on $m_z$ and $m_x$ signs, as sketched in Fig. $\ref{fig:Figure2}$(c). They are associated to Bloch points (with $Q=\pm1$) present at the center of the NdCo layer \cite{NC2020} and their interaction with the closure domain structure of the SML \cite{10.1063/1.4984898}. These four DWs correspond to vortices with $Q=\pm1/2$, or antivortices with $Q=\pm1/2$. In particular, type A-DWs correspond to antivortex textures with $Q=-1/2$, type B-DWs to vortices with $Q=+1/2$, type C-DWs to vortices with $Q=-1/2$ and type D-DWs to antivortices with $Q=+1/2$. Thus, they can be classified into two categories depending on whether the in-plane reversal is head-to-head (Types A and C with $Q=-1/2$) or tail-to-tail (Types B and D with $Q=+1/2$). 

\begin{figure*}[ht]
    \centering
    \includegraphics[width = 0.79\linewidth]{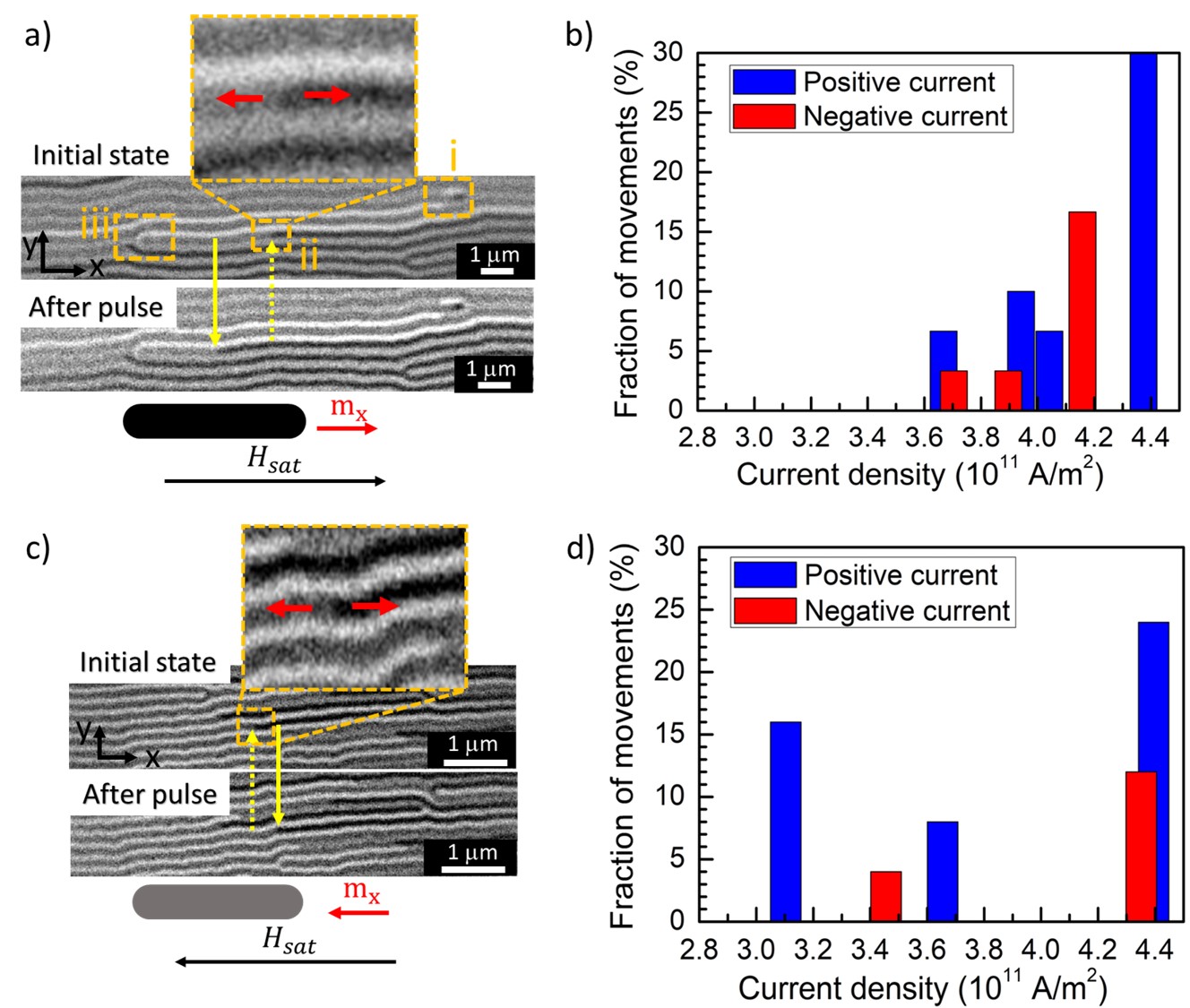}
    \caption{\label{fig:Figure3}Domain wall propagation under current pulses at remanence. (a) MTXM images of the SML before/after a \mbox{$j=4.4\cdot10^{11}$ A/m$^2$} current pulse with the sample at the $[H_{sat}(+),H_{appl}(0)]$ state. Black horizontal arrow indicates the saturation field orientation $H_{sat}(+)$. Orange squares mark regions with different DW  types. Inset shows an enlarged view of the B-DW in region ii with a sketch of m$_x$ configuration (red arrows). Vertical yellow arrows mark the position of the B-DW before (dotted arrow) and after (solid arrow) the current pulse. (b) Statistics of DW propagation events as a function of current density at the $[H_{sat}(+),H_{appl}(0)]$ state: blue/red bars indicate the percentage of moving DWs under current pulses of positive/negative polarity, respectively. (c-d) Same as panels (a-b) with the sample in $[H_{sat}(-),H_{appl}(0)]$ state. Black horizontal arrow indicates the saturation field orientation $H_{sat}(-)$.}
\end{figure*}

\subsection{\label{sec:pulsedI} Guided DW propagation under pulsed currents}

The initial magnetic state was set with the following process: first, the sample was saturated with a positive field parallel to the transport bridge $\mu_0H_x=+200$ mT, then, a small negative field was applied (around half the coercivity, $\mu_0H_x\approx -10$ mT) and, finally, the field amplitude was lowered down to zero until a remnant state with partially reversed in-plane magnetization was reached (denoted as $[H_{sat}(+),H_{appl}(0)]$ state in the following). Different spin textures appear within the SML as seen, for example, in the regions marked in Fig. $\ref{fig:Figure3}$(a): there are three DWs of types B, C and D near a stripe bifurcation at region $i$, a single B-DW at region $ii$ and two B and C DWs  close to the bifurcation in region $iii$.  On average, the density of DWs nucleated within the SML was $\sim 0.15$ $\mu$m$^{-2}$, calculated over the \mbox{39$\times$6.5 $\mu$m$^2$} bridge.
\\
\\

Current pulses with positive and negative polarity were introduced into the transport bridge, progressively increasing  their amplitude until the first DW propagation events were observed above a threshold around \mbox{3.7$\cdot$10$^{11}$ A/m$^2$}. For example, in the MTXM image acquired after a $j=+4.4\cdot10^{11} A/m^2$ pulse (shown in Fig. \ref{fig:Figure3}(a)), the DWs in regions $i$ and $iii$ have remained stationary whereas the B-DW located  in region $ii$ has moved 1.3 $\mu m$ towards the left along the stripe pattern orientation (i.e. $\Delta x = -1.3$ $\mu m$ with a net increase of in-plane magnetization $\Delta m_x>0$ due to the enlargement of the $+|m_x|$ domain). 
Typical propagation distances were in the range 0.8 to 2.3 $\mu m$, always along the direction imposed by the stripe pattern, but without a clear dependence neither on current amplitude nor on DW type.
The main effect of increasing the current amplitude was to induce a larger fraction of propagation events per pulse, with the highest fraction of moving DW ($\sim 30 \%$) obtained for the pulse with the highest amplitude, $j=4.4\cdot 10^{11}$ A/m$^2$. The bar graph of Fig. \ref{fig:Figure3}(b) shows the number of moving DWs for each current pulse over the initial number of existing DWs in the sample. DW propagation statistics in Fig. \ref{fig:Figure3} are very similar for positive (blue bars) and negative (red bars) current pulses. Actually, the propagation sense in each event was not correlated to the sign of the transport current (different from the behavior usually reported in NiFe patterned wires \cite{STT1}). Rather, DW propagation sense was clearly determined by the topological charge of the moving texture. B and D DWs (with $Q=+1/2$) were observed to propagate towards the left (i.e. with $\Delta x <0)$ whereas A and C DWs (with $Q=-1/2$) always moved towards the right (i.e. $\Delta x >0$). A comparison with the sketches of Fig. 2(c) reveals that this combination of opposite propagation senses for DWs with opposite topological charges results in a net increase of in-plane magnetization ($\Delta m_x>0$) in all the observed propagation events, as summarized in the first column of Table \ref{tab:Tab1}.

Figures \ref{fig:Figure3}(c-d) show the results of a similar experiment performed  after bringing the sample close to negative in-plane saturation with $\mu_0H_x=-50$ mT ($[H_{sat}(-),H_{appl}(0)]$ state). Once again, DW propagation is only observed above a threshold current of \mbox{$3.1\cdot 10^{11}$ A/m$^2$} with increasing probability as a function of current amplitude (Fig. \ref{fig:Figure3}(d)). However, in this case, the propagation of the B-DW highlighted in Fig. \ref{fig:Figure3}(c) is towards the right, with $\Delta x \approx +1$ $\mu$m. In fact, the propagation sense of all the different textures in this $[H_{sat}(-),H_{appl}(0)]$ state is reversed in comparison with the results in the $[H_{sat}(+),H_{appl}(0)]$  state (see Table \ref{tab:Tab1}). Thus, in this case, current induced DW propagation events resulted in a net decrease of in-plane magnetization ($\Delta m_x<0$).

\begin{table*}[ht]
\centering
\caption{\label{tab:Tab1} DW propagation sense ($\Delta x$) and corresponding in-plane magnetization change ($\Delta m_x$) as a function of the SML magnetic state. }
\begin{ruledtabular}
\begin{tabular}{cccccccccc}
         &  &  \multicolumn{2}{c}{$[H_{sat}(+),H_{appl}(0)]$}&  \multicolumn{2}{c}{$[H_{sat}(-),H_{appl}(0)]$}&  \multicolumn{2}{c}{$[H_{sat}(+),H_{appl}(-)]$ }&  \multicolumn{2}{c}{$[H_{sat}(-),H_{appl}(+)]$}\\
        \textbf{DW type}&  \textbf{\textit{Q}}& $\Delta x$ & $\Delta m_x$  &  $\Delta x$ &  $\Delta m_x$  &  $\Delta x$ &  $\Delta m_x$  &  $\Delta x$ & $\Delta m_x$  \\
         B,D 
&  +1/2 
&  $-$ 
&  $+$
&  $+$
&  $-$ 
&  $+$
&  $-$ 
&  $-$ 
& $+$
\\
         A,C &  -1/2 &  $+$&  $+$&  
$-$ &  
$-$ 
&  $-$ &  $-$ &  $+$& $+$\\
\end{tabular}
\end{ruledtabular}
\end{table*}

Further pulsed current experiments were performed under constant applied fields, as shown in Fig. 4. First, the sample was taken close to negative in-plane saturation with $\mu_0H_x=-50$ mT and then, positive in-plane fields of increasing magnitude were applied to the system in order to analyze DW propagation events induced by current pulses of constant amplitude $j =4.4\cdot 10^{11}$ A/m$^2$ along a series of consecutive $[H_{sat}(-),H_{appl}(+)]$ states. Figure 4(a) shows a typical propagation event, recorded under a constant field $\mu_0 H_{appl}=+13$ mT, in which a B-DW moves towards the left, with $\Delta x =-16$ $\mu m$. Here, the propagation distance is enhanced by an order of magnitude in comparison with the experiments performed in the remnant state and, more importantly, it occurs in the opposite sense than in the $[H_{sat}(-),H_{appl}(0)]$ state, with a $\Delta m_x>0$ in this case (see Table \ref{tab:Tab1}). Actually, DW propagation is effectively guided by the stripe domain pattern in the range $+10$ mT up to $+30$ mT as the in-plane reversal of the SML proceeds towards positive saturation. In this field range, DW propagation events occur with opposite propagation senses for DWs with opposite $Q$ and always with a net increase of in-plane magnetization. The largest number of DWs and, therefore, the largest number of DW propagation events is observed at $\mu_0H = +13$ mT (with a $\sim 70\%$ probability). As the field increases and the sample approaches positive saturation, the number of DWs decreases and, also, the probability of DW propagation, which is reduced down to $\sim 20\%$ for fields up to $\mu_0H_{appl}\approx 20$ mT. Finally, above 30 mT, no DWs remain in the sample.

Similar results were obtained under negative applied fields ($[H_{sat}(+),H_{appl}(-)]$ state), but with $\Delta m_x <0$ in all cases, as indicated in Table \ref{tab:Tab1}.

\begin{figure}[ht]
    \centering
    \includegraphics[width = 0.8\linewidth]{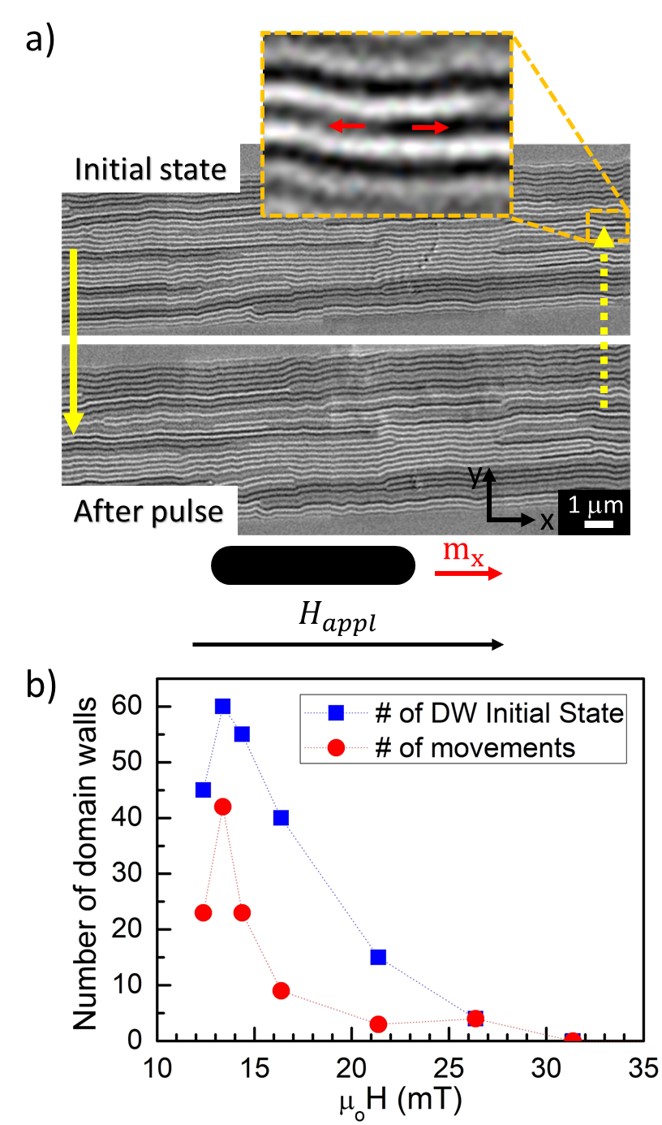}
    \caption{Domain wall propagation under current pulses and an in-plane field ($[H_{sat}(-),H_{appl}(+)]$ state). (a) MTXM images of the SML before/after a $j=4.4\cdot10^{11}$ A/m$^2$ current pulse with $\mu_0H_{appl}=+13$ mT. Inset shows an enlarged view of a B-DW with a sketch of $m_x$ configuration (red arrows). Vertical yellow arrows mark its position before (dotted arrow) and after (solid arrow) the current pulse.  Black horizontal arrow indicates the applied field orientation $H_{appl}(+)$. (b) Statistics of DW propagation as a function of $\mu_0H_{appl}$. Blue dots: number of DWs at the  initial state. Red dots: number of propagation events after a $j=4.4\cdot10^{11}$ A/m$^2$ current pulse.}
    \label{fig:Figure4}
\end{figure}

Then, as summarized in Table \ref{tab:Tab1}, DWs move guided by the direction of the stripe domains with a constant sign of $\Delta m_x$ in each magnetic state that is independent of current polarity. $\Delta m_x$ can be tuned either by applied fields or by magnetic history in the remnant state. 

These facts suggest that DW propagation is a thermally activated phenomenon with a negligible contribution from spin transfer torque effects \cite{STT}. The observed threshold current density is thus an effect of the threshold temperature needed for DW depinning. Once a DW is depinned by the effect of the current pulse, if a constant field is present in the system, it will move in order to enlarge the domain favored by Zeeman energy until it is retrapped at another pinning site. The interaction with the HML stripe pattern hinders the expansion of the reversed domain in the Py layer transverse to the stripes, forcing the guided motion of DWs along them. DW propagation sense will correspond to $\Delta m_x$ following the applied field sign, as observed in Table \ref{tab:Tab1} and similar to the reported behavior of patterned permalloy nanowires under constant applied fields \cite{APLNIFE}. In the remnant state, the correlation between $\Delta m_x$ and the sign of the last saturating field ($H_{sat}$) observed in Table \ref{tab:Tab1} points to the presence of an effective bias field in the system, which keeps memory of the magnetic state in the HML. 

\subsection{\label{sec:Micro} Micromagnetic simulations}

\begin{figure*}[htb]
    \centering
    \includegraphics[width = 0.8\linewidth]{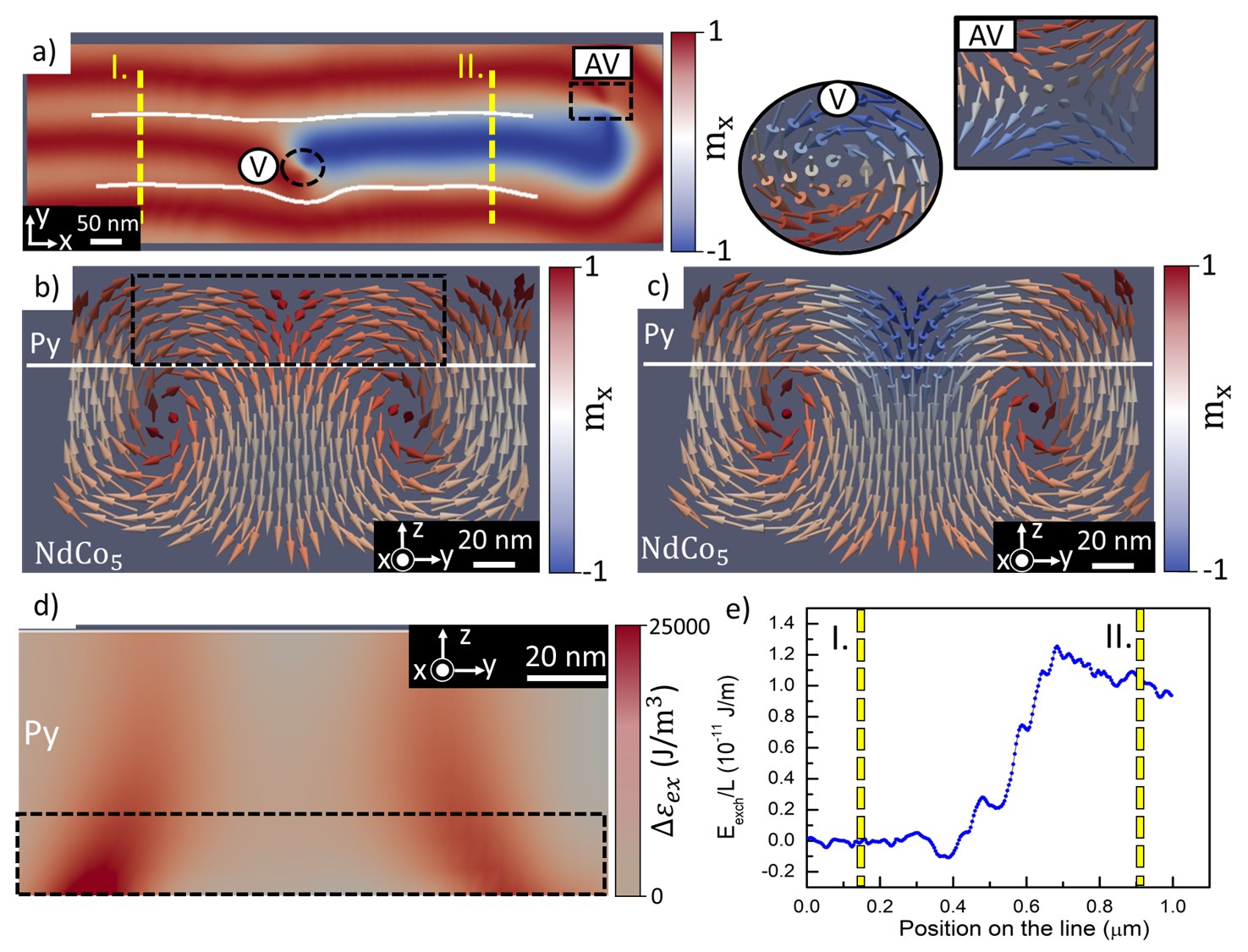}
    \caption{ \label{fig:Figure5}(a) Micromagnetic simulations of a 80 nm NdCo$_5$/40 nm Py bilayer in the $[H_{sat}(+),H_{appl}(0)]$ state. A linear $-|m_x|$ reversed domain is observed at the top of the sample. Dotted circle marks a C-DW containing vortex (V) texture. Dotted rectangle marks an A-DW with an antivortex texture near the bifurcation core. White lines mark the Neel walls that guide the propagating C-DW. (b-c) Micromagnetic configuration of the cross-sections I\&II marked by yellow dotted lines in (a). Note that the $+|m_x|$ orientation set at saturation is stored within the Bloch walls between up/down domains in the NdCo$_5$ layer. (d) Spatial distribution of the difference in exchange energy density $\Delta \varepsilon_{ex}$ between the states in cross sections I\&II, calculated within the dotted rectangle in (b). Note that within the Py layer the highest contribution occurs in the first 12 nm adjacent to the NdCo$_5$ layer. (e) Spatial dependence of exchange energy difference per unit length $\Delta E_{ex}/L$ (integrated within the dotted $\Delta y \times \Delta z = 140$ nm$\times$ 12 nm region in (d)). Note the $\approx 10^{-11}$ J/m increment at the $-|m_x|$ domain. Dotted yellow lines mark the location of cross sections I\&II.
    }
\end{figure*}

Micromagnetic simulations of our HML/SML bilayer confirm that the DW guiding effect and the asymmetries in DW propagation sense observed at remanence are linked to the configuration of the underlying HML. Figure $\ref{fig:Figure5}$ shows the simulated magnetic configuration of a \mbox{80 nm} NdCo$_5$/40 nm Py bilayer in the $[H_{sat}(+),H_{appl}(0)]$ state. A linear reversed $-|m_x|$ domain appears at the top Py layer (see Fig. $\ref{fig:Figure5}$(a) with blue/red $m_x$ contrast). This domain starts at the core of a bifurcation within the stripe pattern and runs along a $-|m_z|$ stripe until it ends at a C-DW (see dotted circle). A zoom into its detailed configuration shows that this C-DW is indeed a twisted vortex spin texture with negative out-of-plane polarity (and $Q=-1/2$).  Another spin texture can be seen near the bifurcation core (see dotted square) that corresponds to an A-DW in the form of an antivortex with positive out-of-plane polarity (see detailed V/AV configurations in the insets of Fig. \ref{fig:Figure5}(a)). 

Figures \ref{fig:Figure5}(b) and \ref{fig:Figure5}(c) show the micromagnetic configuration at both sides of the C-DW (cross sections I\&II indicated by the dashed yellow lines in Fig. \ref{fig:Figure5}(a)). In both cases, there is an up-down alternation of the magnetization within the NdCo$_5$ layer, corresponding to the stripe domain pattern. $\pm|m_z|$ domains are separated by Bloch domain walls with positive $m_x$, following the orientation established during the saturation process with $H_{sat}(+)$. The Py layer hosts the closure domain structure of the NdCo$_5$ stripe pattern, characterized by $m_x$ elongated domains separated by Neel walls that run on top of the Bloch DWs of the NdCo$_5$ layer. The $-|m_x|$ reversed domain can be observed in Fig. \ref{fig:Figure5}(c). It is confined to the 40 nm thick Py layer and lies on top of a $-|m_z|$ domain. It is bounded by two Neel walls which define the \textit{propagation lane} for the C-DW, indicated by white lines in Fig. \ref{fig:Figure5}(a). Any lateral displacement beyond them would imply crossing into a $+|m_z|$ stripe yielding an unfavorable micromagnetic configuration for the vortex texture. Thus, longitudinal displacements along the stripe with matching $m_z$ sign are favored, resulting in the guided propagation of spin textures. 

When the C-DW propagates towards the left (i.e. $\Delta x<0$), the length of the reversed $-|m_x|$ domain increases, modifying the exchange energy in the system due to the antiparallel alignment with the $+|m_x|$ orientation located at the Bloch DW core in the NdCo$_5$ layer. 
A map of $\Delta\varepsilon_{ex}$, the difference in the exchange energy density between the states at cross sections I\&II, displayed in Fig. \ref{fig:Figure5}(d), shows that this additional exchange energy contribution occurs within the domain walls that bound the reversed $-|m_x|$ domain. $\Delta\varepsilon_{ex}$ is concentrated close to the Py/NdCo$_5$ interface, with 45$\%$ of the total $\Delta\varepsilon_{ex}$ occurring within the first 12 nm of the Py layer. Thus, we have estimated the contribution to the exchange energy of the system per unit length of the reversed domain, $\Delta E_{ex}/L$, from the integral of $\Delta\varepsilon_{ex}$ across this 12 nm interfacial layer (marked with a black dotted square in Fig. \ref{fig:Figure5}(d)). 
A plot of the spatial dependence of $\Delta E_{ex}/L$ along the stripe domain shows a step-like increase upon crossing the C-DW position, with an increment \mbox{$\Delta E_{ex}/L\approx1.2\cdot 10^{-11}$ J/m} at the reversed $-|m_x|$ domain with respect to the region with uniform $+|m_x|$  orientation across the thickness. When a spin texture is depinned by a current pulse and is free to propagate along its stripe, the system would tend to lower the energy reducing the length of the reversed domain. Here, it would favor the C-DW propagation towards the right in the micromagnetic simulation ($\Delta m_x>0$), in agreement with the observations in Table \ref{tab:Tab1}. Thus, this $\Delta E_{ex}/L$ contribution acts as a magnetic exchange spring controlling the propagation sense. This spring is related to the $m_x$ sign at the NdCo$_5$ layer, which is determined by the saturation field used to prepare the initial state. Then, the Bloch DWs at the center of the NdCo$_5$ layer store a magnetic memory that favors a specific $m_x$ sense in the top layer of Py. The equivalent bias field can be estimated as $\Delta E_{ex}/L = \frac{1}{2}\mu_0  H_{bias}M_s\Delta y\cdot \tau$ \cite{Mumax} with \mbox{$\tau$ = 40nm}, the Py layer thickness, and \mbox{$\Delta y$ = 140nm}, the width of the reversed domain. The equivalent bias field is \mbox{$\mu_0 H_{bias} \approx$ 10 mT}, about half the value of the coercive field of the SML/HML bilayer.

\section{\label{sec:Conclusions} Conclusions}

Vortex/antivortex propagation has been observed in pulsed current experiments performed by MTXM in HML/SML bilayers. Spin textures move over several $\mu$m distances along the stripe domain pattern of the HML which acts as a natural racetrack. Propagation sense does not depend on the sign of the applied current, indicating a current induced thermally activated depinning of the spin textures above a $3\cdot10^{11}$ A/m$^2$ current density threshold. Opposite propagation senses are observed depending on the sign of the topological charge of each spin texture, resulting in a direct correlation between $\Delta m_x$ and applied field signs. In the remnant state, the propagation sense of spin textures depends on the sign of the last saturating field, indicative of the presence of a bias field in the system. Micromagnetic simulations reveal that the saturation process establishes both the stripe pattern orientation (i.e. the path for the natural racetrack) and the $m_x$ sense at the cores of Bloch walls within the NdCo$_5$ layer, which remains as the field is removed. Thus, the NdCo$_5$ layer stores in its center a magnetic memory that favors spin texture propagation with a specific $\Delta m_x$ sign in the top Py layer. When a spin texture overcomes local pinning by the effect of a current pulse, exchange interactions at the NdCo$_5$/Py interface act as a spring that moves it in the sense determined by the magnetic memory of the NdCo$_5$, with a characteristic bias field of the order of $ \mu_0 H_{bias} \approx$ 10 mT.

Combining a hard magnetic material, with a parallel stripe domain configuration, and a soft magnetic layer, has allowed us to obtain a system that establishes predefined paths for the propagation of spin textures under the effect of pulsed currents. The propagation sense can be controlled taking into account the HML magnetic history and the bias field. Thus, further progress in the development of reconfigurable racetracks for topological spin textures can be expected by tailoring the exchange coupling between the HML and the SML in order to achieve a fine tuning of this bias field.

\acknowledgments
V. V. F. and A. E. H.-A acknowledge the support from
the Severo Ochoa Predoctoral Fellowship Program (nos.PA-22-BP21-124, PA-23-BP22-093, respectively)
The ALBA Synchrotron is funded by the Ministry of Research and Innovation of Spain, by the Generalitat de Catalunya and by European FEDER funds. This work has been supported by Spanish MCIN/AEI/10.13039/501100011033/FEDER,UE under grant PID2022-136784NB and by the ALBA in house research program. 

\bibliography{apssamp}

\begin{thebibliography}{38}%
\makeatletter
\providecommand \@ifxundefined [1]{%
 \@ifx{#1\undefined}
}%
\providecommand \@ifnum [1]{%
 \ifnum #1\expandafter \@firstoftwo
 \else \expandafter \@secondoftwo
 \fi
}%
\providecommand \@ifx [1]{%
 \ifx #1\expandafter \@firstoftwo
 \else \expandafter \@secondoftwo
 \fi
}%
\providecommand \natexlab [1]{#1}%
\providecommand \enquote  [1]{``#1''}%
\providecommand \bibnamefont  [1]{#1}%
\providecommand \bibfnamefont [1]{#1}%
\providecommand \citenamefont [1]{#1}%
\providecommand \href@noop [0]{\@secondoftwo}%
\providecommand \href [0]{\begingroup \@sanitize@url \@href}%
\providecommand \@href[1]{\@@startlink{#1}\@@href}%
\providecommand \@@href[1]{\endgroup#1\@@endlink}%
\providecommand \@sanitize@url [0]{\catcode `\\12\catcode `\$12\catcode `\&12\catcode `\#12\catcode `\^12\catcode `\_12\catcode `\%12\relax}%
\providecommand \@@startlink[1]{}%
\providecommand \@@endlink[0]{}%
\providecommand \url  [0]{\begingroup\@sanitize@url \@url }%
\providecommand \@url [1]{\endgroup\@href {#1}{\urlprefix }}%
\providecommand \urlprefix  [0]{URL }%
\providecommand \Eprint [0]{\href }%
\providecommand \doibase [0]{https://doi.org/}%
\providecommand \selectlanguage [0]{\@gobble}%
\providecommand \bibinfo  [0]{\@secondoftwo}%
\providecommand \bibfield  [0]{\@secondoftwo}%
\providecommand \translation [1]{[#1]}%
\providecommand \BibitemOpen [0]{}%
\providecommand \bibitemStop [0]{}%
\providecommand \bibitemNoStop [0]{.\EOS\space}%
\providecommand \EOS [0]{\spacefactor3000\relax}%
\providecommand \BibitemShut  [1]{\csname bibitem#1\endcsname}%
\let\auto@bib@innerbib\@empty
\bibitem [{\citenamefont {Bläsing}\ \emph {et~al.}(2020)\citenamefont {Bläsing}, \citenamefont {Khan}, \citenamefont {Filippou}, \citenamefont {Garg}, \citenamefont {Hameed}, \citenamefont {Castrillon},\ and\ \citenamefont {Parkin}}]{Parkin}%
  \BibitemOpen
  \bibfield  {author} {\bibinfo {author} {\bibfnamefont {R.}~\bibnamefont {Bläsing}}, \bibinfo {author} {\bibfnamefont {A.~A.}\ \bibnamefont {Khan}}, \bibinfo {author} {\bibfnamefont {P.~C.}\ \bibnamefont {Filippou}}, \bibinfo {author} {\bibfnamefont {C.}~\bibnamefont {Garg}}, \bibinfo {author} {\bibfnamefont {F.}~\bibnamefont {Hameed}}, \bibinfo {author} {\bibfnamefont {J.}~\bibnamefont {Castrillon}},\ and\ \bibinfo {author} {\bibfnamefont {S.~S.~P.}\ \bibnamefont {Parkin}},\ }\href@noop {} {\bibfield  {journal} {\bibinfo  {journal} {Proceedings of the IEEE}\ }\textbf {\bibinfo {volume} {108}},\ \bibinfo {pages} {1303} (\bibinfo {year} {2020})}\BibitemShut {NoStop}%
\bibitem [{\citenamefont {Sanz-Hernández}\ \emph {et~al.}(2021)\citenamefont {Sanz-Hernández}, \citenamefont {Massouras}, \citenamefont {Reyren}, \citenamefont {Rougemaille}, \citenamefont {Schánilec}, \citenamefont {Bouzehouane}, \citenamefont {Hehn}, \citenamefont {Canals}, \citenamefont {Querlioz}, \citenamefont {Grollier}, \citenamefont {Montaigne},\ and\ \citenamefont {Lacour}}]{Dedalo}%
  \BibitemOpen
  \bibfield  {author} {\bibinfo {author} {\bibfnamefont {D.}~\bibnamefont {Sanz-Hernández}}, \bibinfo {author} {\bibfnamefont {M.}~\bibnamefont {Massouras}}, \bibinfo {author} {\bibfnamefont {N.}~\bibnamefont {Reyren}}, \bibinfo {author} {\bibfnamefont {N.}~\bibnamefont {Rougemaille}}, \bibinfo {author} {\bibfnamefont {V.}~\bibnamefont {Schánilec}}, \bibinfo {author} {\bibfnamefont {K.}~\bibnamefont {Bouzehouane}}, \bibinfo {author} {\bibfnamefont {M.}~\bibnamefont {Hehn}}, \bibinfo {author} {\bibfnamefont {B.}~\bibnamefont {Canals}}, \bibinfo {author} {\bibfnamefont {D.}~\bibnamefont {Querlioz}}, \bibinfo {author} {\bibfnamefont {J.}~\bibnamefont {Grollier}}, \bibinfo {author} {\bibfnamefont {F.}~\bibnamefont {Montaigne}},\ and\ \bibinfo {author} {\bibfnamefont {D.}~\bibnamefont {Lacour}},\ }\href@noop {} {\bibfield  {journal} {\bibinfo  {journal} {Advanced Materials}\ }\textbf {\bibinfo {volume} {33}},\ \bibinfo {pages} {2008135} (\bibinfo {year} {2021})}\BibitemShut {NoStop}%
\bibitem [{\citenamefont {Zou}\ \emph {et~al.}(2023)\citenamefont {Zou}, \citenamefont {Bosco}, \citenamefont {Pal}, \citenamefont {Parkin}, \citenamefont {Klinovaja},\ and\ \citenamefont {Loss}}]{quantum}%
  \BibitemOpen
  \bibfield  {author} {\bibinfo {author} {\bibfnamefont {J.}~\bibnamefont {Zou}}, \bibinfo {author} {\bibfnamefont {S.}~\bibnamefont {Bosco}}, \bibinfo {author} {\bibfnamefont {B.}~\bibnamefont {Pal}}, \bibinfo {author} {\bibfnamefont {S.~S.~P.}\ \bibnamefont {Parkin}}, \bibinfo {author} {\bibfnamefont {J.}~\bibnamefont {Klinovaja}},\ and\ \bibinfo {author} {\bibfnamefont {D.}~\bibnamefont {Loss}},\ }\href {https://doi.org/10.1103/PhysRevResearch.5.033166} {\bibfield  {journal} {\bibinfo  {journal} {Phys. Rev. Res.}\ }\textbf {\bibinfo {volume} {5}},\ \bibinfo {pages} {033166} (\bibinfo {year} {2023})}\BibitemShut {NoStop}%
\bibitem [{\citenamefont {Zhang}\ \emph {et~al.}(2012)\citenamefont {Zhang}, \citenamefont {Zhao}, \citenamefont {Ravelosona}, \citenamefont {Klein}, \citenamefont {Kim},\ and\ \citenamefont {Chappert}}]{planar}%
  \BibitemOpen
  \bibfield  {author} {\bibinfo {author} {\bibfnamefont {Y.}~\bibnamefont {Zhang}}, \bibinfo {author} {\bibfnamefont {W.~S.}\ \bibnamefont {Zhao}}, \bibinfo {author} {\bibfnamefont {D.}~\bibnamefont {Ravelosona}}, \bibinfo {author} {\bibfnamefont {J.-O.}\ \bibnamefont {Klein}}, \bibinfo {author} {\bibfnamefont {J.~V.}\ \bibnamefont {Kim}},\ and\ \bibinfo {author} {\bibfnamefont {C.}~\bibnamefont {Chappert}},\ }\href@noop {} {\bibfield  {journal} {\bibinfo  {journal} {Journal of Applied Physics}\ }\textbf {\bibinfo {volume} {111}},\ \bibinfo {pages} {093925} (\bibinfo {year} {2012})}\BibitemShut {NoStop}%
\bibitem [{\citenamefont {Kumar}\ \emph {et~al.}(2022)\citenamefont {Kumar}, \citenamefont {Jin}, \citenamefont {Sbiaa}, \citenamefont {Kl{\"a}ui}, \citenamefont {Bedanta}, \citenamefont {Fukami}, \citenamefont {Ravelosona}, \citenamefont {Yang}, \citenamefont {Liu},\ and\ \citenamefont {Piramanayagam}}]{DWracetrack}%
  \BibitemOpen
  \bibfield  {author} {\bibinfo {author} {\bibfnamefont {D.}~\bibnamefont {Kumar}}, \bibinfo {author} {\bibfnamefont {T.}~\bibnamefont {Jin}}, \bibinfo {author} {\bibfnamefont {R.}~\bibnamefont {Sbiaa}}, \bibinfo {author} {\bibfnamefont {M.}~\bibnamefont {Kl{\"a}ui}}, \bibinfo {author} {\bibfnamefont {S.}~\bibnamefont {Bedanta}}, \bibinfo {author} {\bibfnamefont {S.}~\bibnamefont {Fukami}}, \bibinfo {author} {\bibfnamefont {D.}~\bibnamefont {Ravelosona}}, \bibinfo {author} {\bibfnamefont {S.}~\bibnamefont {Yang}}, \bibinfo {author} {\bibfnamefont {X.}~\bibnamefont {Liu}},\ and\ \bibinfo {author} {\bibfnamefont {S.}~\bibnamefont {Piramanayagam}},\ }\href@noop {} {\bibfield  {journal} {\bibinfo  {journal} {Physics Reports}\ }\textbf {\bibinfo {volume} {958}},\ \bibinfo {pages} {1} (\bibinfo {year} {2022})}\BibitemShut {NoStop}%
\bibitem [{\citenamefont {He}\ \emph {et~al.}(2023)\citenamefont {He}, \citenamefont {Tomasello}, \citenamefont {Luo}, \citenamefont {Zhang}, \citenamefont {Nie}, \citenamefont {Carpentieri}, \citenamefont {Han}, \citenamefont {Finocchio},\ and\ \citenamefont {Yu}}]{skyrmion}%
  \BibitemOpen
  \bibfield  {author} {\bibinfo {author} {\bibfnamefont {B.}~\bibnamefont {He}}, \bibinfo {author} {\bibfnamefont {R.}~\bibnamefont {Tomasello}}, \bibinfo {author} {\bibfnamefont {X.}~\bibnamefont {Luo}}, \bibinfo {author} {\bibfnamefont {R.}~\bibnamefont {Zhang}}, \bibinfo {author} {\bibfnamefont {Z.}~\bibnamefont {Nie}}, \bibinfo {author} {\bibfnamefont {M.}~\bibnamefont {Carpentieri}}, \bibinfo {author} {\bibfnamefont {X.}~\bibnamefont {Han}}, \bibinfo {author} {\bibfnamefont {G.}~\bibnamefont {Finocchio}},\ and\ \bibinfo {author} {\bibfnamefont {G.}~\bibnamefont {Yu}},\ }\href {https://doi.org/10.1021/acs.nanolett.3c02978} {\bibfield  {journal} {\bibinfo  {journal} {Nano Letters}\ }\textbf {\bibinfo {volume} {23}},\ \bibinfo {pages} {9482} (\bibinfo {year} {2023})}\BibitemShut {NoStop}%
\bibitem [{\citenamefont {Fedorov}\ \emph {et~al.}(2024)\citenamefont {Fedorov}, \citenamefont {Soldatov}, \citenamefont {Neu}, \citenamefont {Schäfer}, \citenamefont {Schmidt},\ and\ \citenamefont {Karnaushenko}}]{Schafer}%
  \BibitemOpen
  \bibfield  {author} {\bibinfo {author} {\bibfnamefont {P.}~\bibnamefont {Fedorov}}, \bibinfo {author} {\bibfnamefont {I.}~\bibnamefont {Soldatov}}, \bibinfo {author} {\bibfnamefont {V.}~\bibnamefont {Neu}}, \bibinfo {author} {\bibfnamefont {R.}~\bibnamefont {Schäfer}}, \bibinfo {author} {\bibfnamefont {O.~G.}\ \bibnamefont {Schmidt}},\ and\ \bibinfo {author} {\bibfnamefont {D.}~\bibnamefont {Karnaushenko}},\ }\href {https://doi.org/10.1038/s41467-024-46185-z} {\bibfield  {journal} {\bibinfo  {journal} {Nature Communications}\ }\textbf {\bibinfo {volume} {15}},\ \bibinfo {pages} {2048} (\bibinfo {year} {2024})}\BibitemShut {NoStop}%
\bibitem [{\citenamefont {Hayashi}\ \emph {et~al.}(2006)\citenamefont {Hayashi}, \citenamefont {Thomas}, \citenamefont {Bazaliy}, \citenamefont {Rettner}, \citenamefont {Moriya}, \citenamefont {Jiang},\ and\ \citenamefont {Parkin}}]{STT1}%
  \BibitemOpen
  \bibfield  {author} {\bibinfo {author} {\bibfnamefont {M.}~\bibnamefont {Hayashi}}, \bibinfo {author} {\bibfnamefont {L.}~\bibnamefont {Thomas}}, \bibinfo {author} {\bibfnamefont {Y.~B.}\ \bibnamefont {Bazaliy}}, \bibinfo {author} {\bibfnamefont {C.}~\bibnamefont {Rettner}}, \bibinfo {author} {\bibfnamefont {R.}~\bibnamefont {Moriya}}, \bibinfo {author} {\bibfnamefont {X.}~\bibnamefont {Jiang}},\ and\ \bibinfo {author} {\bibfnamefont {S.~S.~P.}\ \bibnamefont {Parkin}},\ }\href@noop {} {\bibfield  {journal} {\bibinfo  {journal} {Phys. Rev. Lett.}\ }\textbf {\bibinfo {volume} {96}},\ \bibinfo {pages} {197207} (\bibinfo {year} {2006})}\BibitemShut {NoStop}%
\bibitem [{\citenamefont {Yang}\ and\ \citenamefont {Erskine}(2007)}]{STT}%
  \BibitemOpen
  \bibfield  {author} {\bibinfo {author} {\bibfnamefont {S.}~\bibnamefont {Yang}}\ and\ \bibinfo {author} {\bibfnamefont {J.~L.}\ \bibnamefont {Erskine}},\ }\href@noop {} {\bibfield  {journal} {\bibinfo  {journal} {Phys. Rev. B}\ }\textbf {\bibinfo {volume} {75}},\ \bibinfo {pages} {220403} (\bibinfo {year} {2007})}\BibitemShut {NoStop}%
\bibitem [{\citenamefont {Zhang}\ \emph {et~al.}(2016{\natexlab{a}})\citenamefont {Zhang}, \citenamefont {Zhang}, \citenamefont {Hu}, \citenamefont {Nan}, \citenamefont {Zheng}, \citenamefont {Zhang}, \citenamefont {Zhang}, \citenamefont {Vernier}, \citenamefont {Ravelosona},\ and\ \citenamefont {Zhao}}]{SOT}%
  \BibitemOpen
  \bibfield  {author} {\bibinfo {author} {\bibfnamefont {Y.}~\bibnamefont {Zhang}}, \bibinfo {author} {\bibfnamefont {X.}~\bibnamefont {Zhang}}, \bibinfo {author} {\bibfnamefont {J.}~\bibnamefont {Hu}}, \bibinfo {author} {\bibfnamefont {J.}~\bibnamefont {Nan}}, \bibinfo {author} {\bibfnamefont {Z.}~\bibnamefont {Zheng}}, \bibinfo {author} {\bibfnamefont {Z.}~\bibnamefont {Zhang}}, \bibinfo {author} {\bibfnamefont {Y.}~\bibnamefont {Zhang}}, \bibinfo {author} {\bibfnamefont {N.}~\bibnamefont {Vernier}}, \bibinfo {author} {\bibfnamefont {D.}~\bibnamefont {Ravelosona}},\ and\ \bibinfo {author} {\bibfnamefont {W.}~\bibnamefont {Zhao}},\ }\href {https://doi.org/10.1038/srep35062} {\bibfield  {journal} {\bibinfo  {journal} {Scientific Reports}\ }\textbf {\bibinfo {volume} {6}},\ \bibinfo {pages} {35062} (\bibinfo {year} {2016}{\natexlab{a}})}\BibitemShut {NoStop}%
\bibitem [{\citenamefont {Sch\"obitz}\ \emph {et~al.}(2019)\citenamefont {Sch\"obitz}, \citenamefont {De~Riz}, \citenamefont {Martin}, \citenamefont {Bochmann}, \citenamefont {Thirion}, \citenamefont {Vogel}, \citenamefont {Foerster}, \citenamefont {Aballe}, \citenamefont {Mente\ifmmode~\mbox{\c{s}}\else \c{s}\fi{}}, \citenamefont {Locatelli}, \citenamefont {Genuzio}, \citenamefont {Le-Denmat}, \citenamefont {Cagnon}, \citenamefont {Toussaint}, \citenamefont {Gusakova}, \citenamefont {Bachmann},\ and\ \citenamefont {Fruchart}}]{Fruchart}%
  \BibitemOpen
  \bibfield  {author} {\bibinfo {author} {\bibfnamefont {M.}~\bibnamefont {Sch\"obitz}}, \bibinfo {author} {\bibfnamefont {A.}~\bibnamefont {De~Riz}}, \bibinfo {author} {\bibfnamefont {S.}~\bibnamefont {Martin}}, \bibinfo {author} {\bibfnamefont {S.}~\bibnamefont {Bochmann}}, \bibinfo {author} {\bibfnamefont {C.}~\bibnamefont {Thirion}}, \bibinfo {author} {\bibfnamefont {J.}~\bibnamefont {Vogel}}, \bibinfo {author} {\bibfnamefont {M.}~\bibnamefont {Foerster}}, \bibinfo {author} {\bibfnamefont {L.}~\bibnamefont {Aballe}}, \bibinfo {author} {\bibfnamefont {T.~O.}\ \bibnamefont {Mente\ifmmode~\mbox{\c{s}}\else \c{s}\fi{}}}, \bibinfo {author} {\bibfnamefont {A.}~\bibnamefont {Locatelli}}, \bibinfo {author} {\bibfnamefont {F.}~\bibnamefont {Genuzio}}, \bibinfo {author} {\bibfnamefont {S.}~\bibnamefont {Le-Denmat}}, \bibinfo {author} {\bibfnamefont {L.}~\bibnamefont {Cagnon}}, \bibinfo {author} {\bibfnamefont {J.~C.}\ \bibnamefont {Toussaint}}, \bibinfo {author} {\bibfnamefont {D.}~\bibnamefont {Gusakova}},
  \bibinfo {author} {\bibfnamefont {J.}~\bibnamefont {Bachmann}},\ and\ \bibinfo {author} {\bibfnamefont {O.}~\bibnamefont {Fruchart}},\ }\href@noop {} {\bibfield  {journal} {\bibinfo  {journal} {Phys. Rev. Lett.}\ }\textbf {\bibinfo {volume} {123}},\ \bibinfo {pages} {217201} (\bibinfo {year} {2019})}\BibitemShut {NoStop}%
\bibitem [{\citenamefont {Bran}\ \emph {et~al.}(2023)\citenamefont {Bran}, \citenamefont {Fernandez-Roldan}, \citenamefont {Moreno}, \citenamefont {Fraile~Rodríguez}, \citenamefont {del Real}, \citenamefont {Asenjo}, \citenamefont {Saugar}, \citenamefont {Marqués-Marchán}, \citenamefont {Mohammed}, \citenamefont {Foerster}, \citenamefont {Aballe}, \citenamefont {Kosel}, \citenamefont {Vazquez},\ and\ \citenamefont {Chubykalo-Fesenko}}]{nw}%
  \BibitemOpen
  \bibfield  {author} {\bibinfo {author} {\bibfnamefont {C.}~\bibnamefont {Bran}}, \bibinfo {author} {\bibfnamefont {J.~A.}\ \bibnamefont {Fernandez-Roldan}}, \bibinfo {author} {\bibfnamefont {J.~A.}\ \bibnamefont {Moreno}}, \bibinfo {author} {\bibfnamefont {A.}~\bibnamefont {Fraile~Rodríguez}}, \bibinfo {author} {\bibfnamefont {R.~P.}\ \bibnamefont {del Real}}, \bibinfo {author} {\bibfnamefont {A.}~\bibnamefont {Asenjo}}, \bibinfo {author} {\bibfnamefont {E.}~\bibnamefont {Saugar}}, \bibinfo {author} {\bibfnamefont {J.}~\bibnamefont {Marqués-Marchán}}, \bibinfo {author} {\bibfnamefont {H.}~\bibnamefont {Mohammed}}, \bibinfo {author} {\bibfnamefont {M.}~\bibnamefont {Foerster}}, \bibinfo {author} {\bibfnamefont {L.}~\bibnamefont {Aballe}}, \bibinfo {author} {\bibfnamefont {J.}~\bibnamefont {Kosel}}, \bibinfo {author} {\bibfnamefont {M.}~\bibnamefont {Vazquez}},\ and\ \bibinfo {author} {\bibfnamefont {O.}~\bibnamefont {Chubykalo-Fesenko}},\ }\href@noop {} {\bibfield  {journal} {\bibinfo  {journal}
  {Nanoscale}\ }\textbf {\bibinfo {volume} {15}},\ \bibinfo {pages} {8387} (\bibinfo {year} {2023})}\BibitemShut {NoStop}%
\bibitem [{\citenamefont {Alvaro-Gómez}\ \emph {et~al.}(2024)\citenamefont {Alvaro-Gómez}, \citenamefont {Masseboeuf}, \citenamefont {Mille}, \citenamefont {ález}, \citenamefont {Ruiz-Gómez}, \citenamefont {Toussaint}, \citenamefont {Belkhou}, \citenamefont {Foerster}, \citenamefont {Pereiro}, \citenamefont {Aballe}, \citenamefont {Thirion}, \citenamefont {Gusakova}, \citenamefont {Fruchart},\ and\ \citenamefont {Pérez}}]{Lucas}%
  \BibitemOpen
  \bibfield  {author} {\bibinfo {author} {\bibfnamefont {L.}~\bibnamefont {Alvaro-Gómez}}, \bibinfo {author} {\bibfnamefont {A.}~\bibnamefont {Masseboeuf}}, \bibinfo {author} {\bibfnamefont {N.}~\bibnamefont {Mille}}, \bibinfo {author} {\bibfnamefont {C.~F.-G.}\ \bibnamefont {ález}}, \bibinfo {author} {\bibfnamefont {S.}~\bibnamefont {Ruiz-Gómez}}, \bibinfo {author} {\bibfnamefont {J.~C.}\ \bibnamefont {Toussaint}}, \bibinfo {author} {\bibfnamefont {R.}~\bibnamefont {Belkhou}}, \bibinfo {author} {\bibfnamefont {M.}~\bibnamefont {Foerster}}, \bibinfo {author} {\bibfnamefont {E.}~\bibnamefont {Pereiro}}, \bibinfo {author} {\bibfnamefont {L.}~\bibnamefont {Aballe}}, \bibinfo {author} {\bibfnamefont {C.}~\bibnamefont {Thirion}}, \bibinfo {author} {\bibfnamefont {D.}~\bibnamefont {Gusakova}}, \bibinfo {author} {\bibfnamefont {O.}~\bibnamefont {Fruchart}},\ and\ \bibinfo {author} {\bibfnamefont {L.}~\bibnamefont {Pérez}},\ }\href@noop {} {\bibinfo {title} {Interplay between domain walls and magnetization
  curling induced by chemical modulations in cylindrical nanowires}} (\bibinfo {year} {2024}),\ \Eprint {https://arxiv.org/abs/2405.00652} {arXiv:2405.00652} \BibitemShut {NoStop}%
\bibitem [{\citenamefont {Fangohr}\ \emph {et~al.}(2011)\citenamefont {Fangohr}, \citenamefont {Chernyshenko}, \citenamefont {Franchin}, \citenamefont {Fischbacher},\ and\ \citenamefont {Meier}}]{Joule}%
  \BibitemOpen
  \bibfield  {author} {\bibinfo {author} {\bibfnamefont {H.}~\bibnamefont {Fangohr}}, \bibinfo {author} {\bibfnamefont {D.~S.}\ \bibnamefont {Chernyshenko}}, \bibinfo {author} {\bibfnamefont {M.}~\bibnamefont {Franchin}}, \bibinfo {author} {\bibfnamefont {T.}~\bibnamefont {Fischbacher}},\ and\ \bibinfo {author} {\bibfnamefont {G.}~\bibnamefont {Meier}},\ }\href {https://doi.org/10.1103/PhysRevB.84.054437} {\bibfield  {journal} {\bibinfo  {journal} {Phys. Rev. B}\ }\textbf {\bibinfo {volume} {84}},\ \bibinfo {pages} {054437} (\bibinfo {year} {2011})}\BibitemShut {NoStop}%
\bibitem [{\citenamefont {Skoric}\ \emph {et~al.}(2022)\citenamefont {Skoric}, \citenamefont {Donnelly}, \citenamefont {Hierro-Rodriguez}, \citenamefont {Cascales~Sandoval}, \citenamefont {Ruiz-Gómez}, \citenamefont {Foerster}, \citenamefont {Niño}, \citenamefont {Belkhou}, \citenamefont {Abert}, \citenamefont {Suess},\ and\ \citenamefont {Fernández-Pacheco}}]{ACSNano22}%
  \BibitemOpen
  \bibfield  {author} {\bibinfo {author} {\bibfnamefont {L.}~\bibnamefont {Skoric}}, \bibinfo {author} {\bibfnamefont {C.}~\bibnamefont {Donnelly}}, \bibinfo {author} {\bibfnamefont {A.}~\bibnamefont {Hierro-Rodriguez}}, \bibinfo {author} {\bibfnamefont {M.~A.}\ \bibnamefont {Cascales~Sandoval}}, \bibinfo {author} {\bibfnamefont {S.}~\bibnamefont {Ruiz-Gómez}}, \bibinfo {author} {\bibfnamefont {M.}~\bibnamefont {Foerster}}, \bibinfo {author} {\bibfnamefont {M.~A.}\ \bibnamefont {Niño}}, \bibinfo {author} {\bibfnamefont {R.}~\bibnamefont {Belkhou}}, \bibinfo {author} {\bibfnamefont {C.}~\bibnamefont {Abert}}, \bibinfo {author} {\bibfnamefont {D.}~\bibnamefont {Suess}},\ and\ \bibinfo {author} {\bibfnamefont {A.}~\bibnamefont {Fernández-Pacheco}},\ }\href {https://doi.org/10.1021/acsnano.1c10345} {\bibfield  {journal} {\bibinfo  {journal} {ACS Nano}\ }\textbf {\bibinfo {volume} {16}},\ \bibinfo {pages} {8860} (\bibinfo {year} {2022})},\ \bibinfo {note} {pMID: 35580039}\BibitemShut {NoStop}%
\bibitem [{\citenamefont {Zhang}\ \emph {et~al.}(2016{\natexlab{b}})\citenamefont {Zhang}, \citenamefont {Ezawa},\ and\ \citenamefont {Zhou}}]{SAF}%
  \BibitemOpen
  \bibfield  {author} {\bibinfo {author} {\bibfnamefont {X.}~\bibnamefont {Zhang}}, \bibinfo {author} {\bibfnamefont {M.}~\bibnamefont {Ezawa}},\ and\ \bibinfo {author} {\bibfnamefont {Y.}~\bibnamefont {Zhou}},\ }\href {https://doi.org/10.1103/PhysRevB.94.064406} {\bibfield  {journal} {\bibinfo  {journal} {Phys. Rev. B}\ }\textbf {\bibinfo {volume} {94}},\ \bibinfo {pages} {064406} (\bibinfo {year} {2016}{\natexlab{b}})}\BibitemShut {NoStop}%
\bibitem [{\citenamefont {G\"obel}\ \emph {et~al.}(2019)\citenamefont {G\"obel}, \citenamefont {Mook}, \citenamefont {Henk}, \citenamefont {Mertig},\ and\ \citenamefont {Tretiakov}}]{bimeron}%
  \BibitemOpen
  \bibfield  {author} {\bibinfo {author} {\bibfnamefont {B.}~\bibnamefont {G\"obel}}, \bibinfo {author} {\bibfnamefont {A.}~\bibnamefont {Mook}}, \bibinfo {author} {\bibfnamefont {J.}~\bibnamefont {Henk}}, \bibinfo {author} {\bibfnamefont {I.}~\bibnamefont {Mertig}},\ and\ \bibinfo {author} {\bibfnamefont {O.~A.}\ \bibnamefont {Tretiakov}},\ }\href {https://doi.org/10.1103/PhysRevB.99.060407} {\bibfield  {journal} {\bibinfo  {journal} {Phys. Rev. B}\ }\textbf {\bibinfo {volume} {99}},\ \bibinfo {pages} {060407} (\bibinfo {year} {2019})}\BibitemShut {NoStop}%
\bibitem [{\citenamefont {Ara\'ujo}\ \emph {et~al.}(2020)\citenamefont {Ara\'ujo}, \citenamefont {Lopes}, \citenamefont {Carvalho-Santos}, \citenamefont {Pereira}, \citenamefont {Silva}, \citenamefont {Silva},\ and\ \citenamefont {Altbir}}]{bimeron2}%
  \BibitemOpen
  \bibfield  {author} {\bibinfo {author} {\bibfnamefont {A.~S.}\ \bibnamefont {Ara\'ujo}}, \bibinfo {author} {\bibfnamefont {R.~J.~C.}\ \bibnamefont {Lopes}}, \bibinfo {author} {\bibfnamefont {V.~L.}\ \bibnamefont {Carvalho-Santos}}, \bibinfo {author} {\bibfnamefont {A.~R.}\ \bibnamefont {Pereira}}, \bibinfo {author} {\bibfnamefont {R.~L.}\ \bibnamefont {Silva}}, \bibinfo {author} {\bibfnamefont {R.~C.}\ \bibnamefont {Silva}},\ and\ \bibinfo {author} {\bibfnamefont {D.}~\bibnamefont {Altbir}},\ }\href {https://doi.org/10.1103/PhysRevB.102.104409} {\bibfield  {journal} {\bibinfo  {journal} {Phys. Rev. B}\ }\textbf {\bibinfo {volume} {102}},\ \bibinfo {pages} {104409} (\bibinfo {year} {2020})}\BibitemShut {NoStop}%
\bibitem [{\citenamefont {Navas}\ \emph {et~al.}(2019)\citenamefont {Navas}, \citenamefont {Verba}, \citenamefont {Hierro-Rodriguez}, \citenamefont {Bunyaev}, \citenamefont {Zhou}, \citenamefont {Adeyeye}, \citenamefont {Dobrovolskiy}, \citenamefont {Ivanov}, \citenamefont {Guslienko},\ and\ \citenamefont {Kakazei}}]{dipolar}%
  \BibitemOpen
  \bibfield  {author} {\bibinfo {author} {\bibfnamefont {D.}~\bibnamefont {Navas}}, \bibinfo {author} {\bibfnamefont {R.~V.}\ \bibnamefont {Verba}}, \bibinfo {author} {\bibfnamefont {A.}~\bibnamefont {Hierro-Rodriguez}}, \bibinfo {author} {\bibfnamefont {S.~A.}\ \bibnamefont {Bunyaev}}, \bibinfo {author} {\bibfnamefont {X.}~\bibnamefont {Zhou}}, \bibinfo {author} {\bibfnamefont {A.~O.}\ \bibnamefont {Adeyeye}}, \bibinfo {author} {\bibfnamefont {O.~V.}\ \bibnamefont {Dobrovolskiy}}, \bibinfo {author} {\bibfnamefont {B.~A.}\ \bibnamefont {Ivanov}}, \bibinfo {author} {\bibfnamefont {K.~Y.}\ \bibnamefont {Guslienko}},\ and\ \bibinfo {author} {\bibfnamefont {G.~N.}\ \bibnamefont {Kakazei}},\ }\href@noop {} {\bibfield  {journal} {\bibinfo  {journal} {APL Materials}\ }\textbf {\bibinfo {volume} {7}},\ \bibinfo {pages} {081114} (\bibinfo {year} {2019})}\BibitemShut {NoStop}%
\bibitem [{\citenamefont {Lee}\ \emph {et~al.}(2023)\citenamefont {Lee}, \citenamefont {Jeong}, \citenamefont {Kim},\ and\ \citenamefont {Kim}}]{reconfig1}%
  \BibitemOpen
  \bibfield  {author} {\bibinfo {author} {\bibfnamefont {T.}~\bibnamefont {Lee}}, \bibinfo {author} {\bibfnamefont {S.}~\bibnamefont {Jeong}}, \bibinfo {author} {\bibfnamefont {S.}~\bibnamefont {Kim}},\ and\ \bibinfo {author} {\bibfnamefont {K.-J.}\ \bibnamefont {Kim}},\ }\href {https://doi.org/10.1038/s41598-023-34040-y} {\bibfield  {journal} {\bibinfo  {journal} {Scientific Reports}\ }\textbf {\bibinfo {volume} {13}},\ \bibinfo {pages} {6791} (\bibinfo {year} {2023})}\BibitemShut {NoStop}%
\bibitem [{\citenamefont {Hurst}\ \emph {et~al.}(2017)\citenamefont {Hurst}, \citenamefont {Izaac}, \citenamefont {Altaf}, \citenamefont {Baltz},\ and\ \citenamefont {Metaxas}}]{reconfig2}%
  \BibitemOpen
  \bibfield  {author} {\bibinfo {author} {\bibfnamefont {A.~C.~H.}\ \bibnamefont {Hurst}}, \bibinfo {author} {\bibfnamefont {J.~A.}\ \bibnamefont {Izaac}}, \bibinfo {author} {\bibfnamefont {F.}~\bibnamefont {Altaf}}, \bibinfo {author} {\bibfnamefont {V.}~\bibnamefont {Baltz}},\ and\ \bibinfo {author} {\bibfnamefont {P.~J.}\ \bibnamefont {Metaxas}},\ }\href@noop {} {\bibfield  {journal} {\bibinfo  {journal} {Applied Physics Letters}\ }\textbf {\bibinfo {volume} {110}},\ \bibinfo {pages} {182404} (\bibinfo {year} {2017})}\BibitemShut {NoStop}%
\bibitem [{\citenamefont {Fin}\ \emph {et~al.}(2018)\citenamefont {Fin}, \citenamefont {Silvani}, \citenamefont {Tacchi}, \citenamefont {Marangolo}, \citenamefont {Garnier}, \citenamefont {Eddrief}, \citenamefont {Hepburn}, \citenamefont {Fortuna}, \citenamefont {Rettori}, \citenamefont {Pini},\ and\ \citenamefont {Bisero}}]{Italiano}%
  \BibitemOpen
  \bibfield  {author} {\bibinfo {author} {\bibfnamefont {S.}~\bibnamefont {Fin}}, \bibinfo {author} {\bibfnamefont {R.}~\bibnamefont {Silvani}}, \bibinfo {author} {\bibfnamefont {S.}~\bibnamefont {Tacchi}}, \bibinfo {author} {\bibfnamefont {M.}~\bibnamefont {Marangolo}}, \bibinfo {author} {\bibfnamefont {L.~C.}\ \bibnamefont {Garnier}}, \bibinfo {author} {\bibfnamefont {M.}~\bibnamefont {Eddrief}}, \bibinfo {author} {\bibfnamefont {C.}~\bibnamefont {Hepburn}}, \bibinfo {author} {\bibfnamefont {F.}~\bibnamefont {Fortuna}}, \bibinfo {author} {\bibfnamefont {A.}~\bibnamefont {Rettori}}, \bibinfo {author} {\bibfnamefont {M.~G.}\ \bibnamefont {Pini}},\ and\ \bibinfo {author} {\bibfnamefont {D.}~\bibnamefont {Bisero}},\ }\href@noop {} {\bibfield  {journal} {\bibinfo  {journal} {Scientific Reports}\ }\textbf {\bibinfo {volume} {8}},\ \bibinfo {pages} {9339} (\bibinfo {year} {2018})}\BibitemShut {NoStop}%
\bibitem [{\citenamefont {Luo}\ \emph {et~al.}(2020)\citenamefont {Luo}, \citenamefont {Hrabec}, \citenamefont {Dao}, \citenamefont {Sala}, \citenamefont {Finizio}, \citenamefont {Feng}, \citenamefont {Mayr}, \citenamefont {Raabe}, \citenamefont {Gambardella},\ and\ \citenamefont {Heyderman}}]{DWlogic}%
  \BibitemOpen
  \bibfield  {author} {\bibinfo {author} {\bibfnamefont {Z.}~\bibnamefont {Luo}}, \bibinfo {author} {\bibfnamefont {A.}~\bibnamefont {Hrabec}}, \bibinfo {author} {\bibfnamefont {T.~P.}\ \bibnamefont {Dao}}, \bibinfo {author} {\bibfnamefont {G.}~\bibnamefont {Sala}}, \bibinfo {author} {\bibfnamefont {S.}~\bibnamefont {Finizio}}, \bibinfo {author} {\bibfnamefont {J.}~\bibnamefont {Feng}}, \bibinfo {author} {\bibfnamefont {S.}~\bibnamefont {Mayr}}, \bibinfo {author} {\bibfnamefont {J.}~\bibnamefont {Raabe}}, \bibinfo {author} {\bibfnamefont {P.}~\bibnamefont {Gambardella}},\ and\ \bibinfo {author} {\bibfnamefont {L.~J.}\ \bibnamefont {Heyderman}},\ }\href {https://doi.org/10.1038/s41586-020-2061-y} {\bibfield  {journal} {\bibinfo  {journal} {Nature}\ }\textbf {\bibinfo {volume} {579}},\ \bibinfo {pages} {214} (\bibinfo {year} {2020})}\BibitemShut {NoStop}%
\bibitem [{\citenamefont {Hubert}\ and\ \citenamefont {Schäfer}(1998)}]{huber}%
  \BibitemOpen
  \bibfield  {author} {\bibinfo {author} {\bibfnamefont {A.}~\bibnamefont {Hubert}}\ and\ \bibinfo {author} {\bibfnamefont {R.}~\bibnamefont {Schäfer}},\ }\href@noop {} {\emph {\bibinfo {title} {Magnetic Domains: The Analysis of Magnetic Microstructures}}}\ (\bibinfo  {publisher} {Springer-Verlag, Berlin},\ \bibinfo {year} {1998})\BibitemShut {NoStop}%
\bibitem [{\citenamefont {Hierro-Rodriguez}\ \emph {et~al.}(2013)\citenamefont {Hierro-Rodriguez}, \citenamefont {V\'elez}, \citenamefont {Morales}, \citenamefont {Soriano}, \citenamefont {Rodr\'{\i}guez-Rodr\'{\i}guez}, \citenamefont {\'Alvarez-Prado}, \citenamefont {Mart\'{\i}n},\ and\ \citenamefont {Alameda}}]{PRB2013}%
  \BibitemOpen
  \bibfield  {author} {\bibinfo {author} {\bibfnamefont {A.}~\bibnamefont {Hierro-Rodriguez}}, \bibinfo {author} {\bibfnamefont {M.}~\bibnamefont {V\'elez}}, \bibinfo {author} {\bibfnamefont {R.}~\bibnamefont {Morales}}, \bibinfo {author} {\bibfnamefont {N.}~\bibnamefont {Soriano}}, \bibinfo {author} {\bibfnamefont {G.}~\bibnamefont {Rodr\'{\i}guez-Rodr\'{\i}guez}}, \bibinfo {author} {\bibfnamefont {L.~M.}\ \bibnamefont {\'Alvarez-Prado}}, \bibinfo {author} {\bibfnamefont {J.~I.}\ \bibnamefont {Mart\'{\i}n}},\ and\ \bibinfo {author} {\bibfnamefont {J.~M.}\ \bibnamefont {Alameda}},\ }\href {https://doi.org/10.1103/PhysRevB.88.174411} {\bibfield  {journal} {\bibinfo  {journal} {Phys. Rev. B}\ }\textbf {\bibinfo {volume} {88}},\ \bibinfo {pages} {174411} (\bibinfo {year} {2013})}\BibitemShut {NoStop}%
\bibitem [{\citenamefont {Fujiwara}\ \emph {et~al.}(1964)\citenamefont {Fujiwara}, \citenamefont {Sugita},\ and\ \citenamefont {Saito}}]{OriginRotAnis}%
  \BibitemOpen
  \bibfield  {author} {\bibinfo {author} {\bibfnamefont {H.}~\bibnamefont {Fujiwara}}, \bibinfo {author} {\bibfnamefont {Y.}~\bibnamefont {Sugita}},\ and\ \bibinfo {author} {\bibfnamefont {N.}~\bibnamefont {Saito}},\ }\href {https://doi.org/10.1063/1.1753938} {\bibfield  {journal} {\bibinfo  {journal} {Applied Physics Letters}\ }\textbf {\bibinfo {volume} {4}},\ \bibinfo {pages} {199} (\bibinfo {year} {1964})}\BibitemShut {NoStop}%
\bibitem [{\citenamefont {Hierro-Rodriguez}\ \emph {et~al.}(2012)\citenamefont {Hierro-Rodriguez}, \citenamefont {Cid}, \citenamefont {V\'elez}, \citenamefont {Rodriguez-Rodriguez}, \citenamefont {Mart\'{\i}n}, \citenamefont {\'Alvarez-Prado},\ and\ \citenamefont {Alameda}}]{PRL2012}%
  \BibitemOpen
  \bibfield  {author} {\bibinfo {author} {\bibfnamefont {A.}~\bibnamefont {Hierro-Rodriguez}}, \bibinfo {author} {\bibfnamefont {R.}~\bibnamefont {Cid}}, \bibinfo {author} {\bibfnamefont {M.}~\bibnamefont {V\'elez}}, \bibinfo {author} {\bibfnamefont {G.}~\bibnamefont {Rodriguez-Rodriguez}}, \bibinfo {author} {\bibfnamefont {J.~I.}\ \bibnamefont {Mart\'{\i}n}}, \bibinfo {author} {\bibfnamefont {L.~M.}\ \bibnamefont {\'Alvarez-Prado}},\ and\ \bibinfo {author} {\bibfnamefont {J.~M.}\ \bibnamefont {Alameda}},\ }\href {https://doi.org/10.1103/PhysRevLett.109.117202} {\bibfield  {journal} {\bibinfo  {journal} {Phys. Rev. Lett.}\ }\textbf {\bibinfo {volume} {109}},\ \bibinfo {pages} {117202} (\bibinfo {year} {2012})}\BibitemShut {NoStop}%
\bibitem [{\citenamefont {Hierro-Rodriguez}\ \emph {et~al.}(2014)\citenamefont {Hierro-Rodriguez}, \citenamefont {Teixeira}, \citenamefont {Vélez}, \citenamefont {Alvarez-Prado}, \citenamefont {Martín},\ and\ \citenamefont {Alameda}}]{APL2014}%
  \BibitemOpen
  \bibfield  {author} {\bibinfo {author} {\bibfnamefont {A.}~\bibnamefont {Hierro-Rodriguez}}, \bibinfo {author} {\bibfnamefont {J.~M.}\ \bibnamefont {Teixeira}}, \bibinfo {author} {\bibfnamefont {M.}~\bibnamefont {Vélez}}, \bibinfo {author} {\bibfnamefont {L.~M.}\ \bibnamefont {Alvarez-Prado}}, \bibinfo {author} {\bibfnamefont {J.~I.}\ \bibnamefont {Martín}},\ and\ \bibinfo {author} {\bibfnamefont {J.~M.}\ \bibnamefont {Alameda}},\ }\href@noop {} {\bibfield  {journal} {\bibinfo  {journal} {Applied Physics Letters}\ }\textbf {\bibinfo {volume} {105}},\ \bibinfo {pages} {102412} (\bibinfo {year} {2014})}\BibitemShut {NoStop}%
\bibitem [{\citenamefont {Blanco-Roldán}\ \emph {et~al.}(2015)\citenamefont {Blanco-Roldán}, \citenamefont {Quirós}, \citenamefont {Sorrentino}, \citenamefont {Hierro-Rodríguez}, \citenamefont {Álvarez Prado}, \citenamefont {Valcárcel}, \citenamefont {Duch}, \citenamefont {Torras}, \citenamefont {Esteve}, \citenamefont {Martín}, \citenamefont {Vélez}, \citenamefont {Alameda}, \citenamefont {Pereiro},\ and\ \citenamefont {Ferrer}}]{NC2015}%
  \BibitemOpen
  \bibfield  {author} {\bibinfo {author} {\bibfnamefont {C.}~\bibnamefont {Blanco-Roldán}}, \bibinfo {author} {\bibfnamefont {C.}~\bibnamefont {Quirós}}, \bibinfo {author} {\bibfnamefont {A.}~\bibnamefont {Sorrentino}}, \bibinfo {author} {\bibfnamefont {A.}~\bibnamefont {Hierro-Rodríguez}}, \bibinfo {author} {\bibfnamefont {L.~M.}\ \bibnamefont {Álvarez Prado}}, \bibinfo {author} {\bibfnamefont {R.}~\bibnamefont {Valcárcel}}, \bibinfo {author} {\bibfnamefont {M.}~\bibnamefont {Duch}}, \bibinfo {author} {\bibfnamefont {N.}~\bibnamefont {Torras}}, \bibinfo {author} {\bibfnamefont {J.}~\bibnamefont {Esteve}}, \bibinfo {author} {\bibfnamefont {J.~I.}\ \bibnamefont {Martín}}, \bibinfo {author} {\bibfnamefont {M.}~\bibnamefont {Vélez}}, \bibinfo {author} {\bibfnamefont {J.~M.}\ \bibnamefont {Alameda}}, \bibinfo {author} {\bibfnamefont {E.}~\bibnamefont {Pereiro}},\ and\ \bibinfo {author} {\bibfnamefont {S.}~\bibnamefont {Ferrer}},\ }\href@noop {} {\bibfield  {journal} {\bibinfo  {journal} {Nat. Commun.}\
  }\textbf {\bibinfo {volume} {6}},\ \bibinfo {pages} {8196} (\bibinfo {year} {2015})}\BibitemShut {NoStop}%
\bibitem [{\citenamefont {Hierro-Rodriguez}\ \emph {et~al.}(2017{\natexlab{a}})\citenamefont {Hierro-Rodriguez}, \citenamefont {Quir\'os}, \citenamefont {Sorrentino}, \citenamefont {Blanco-Rold\'an}, \citenamefont {Alvarez-Prado}, \citenamefont {Mart\'{\i}n}, \citenamefont {Alameda}, \citenamefont {Pereiro}, \citenamefont {V\'elez},\ and\ \citenamefont {Ferrer}}]{PRB2017}%
  \BibitemOpen
  \bibfield  {author} {\bibinfo {author} {\bibfnamefont {A.}~\bibnamefont {Hierro-Rodriguez}}, \bibinfo {author} {\bibfnamefont {C.}~\bibnamefont {Quir\'os}}, \bibinfo {author} {\bibfnamefont {A.}~\bibnamefont {Sorrentino}}, \bibinfo {author} {\bibfnamefont {C.}~\bibnamefont {Blanco-Rold\'an}}, \bibinfo {author} {\bibfnamefont {L.~M.}\ \bibnamefont {Alvarez-Prado}}, \bibinfo {author} {\bibfnamefont {J.~I.}\ \bibnamefont {Mart\'{\i}n}}, \bibinfo {author} {\bibfnamefont {J.~M.}\ \bibnamefont {Alameda}}, \bibinfo {author} {\bibfnamefont {E.}~\bibnamefont {Pereiro}}, \bibinfo {author} {\bibfnamefont {M.}~\bibnamefont {V\'elez}},\ and\ \bibinfo {author} {\bibfnamefont {S.}~\bibnamefont {Ferrer}},\ }\href@noop {} {\bibfield  {journal} {\bibinfo  {journal} {Phys. Rev. B}\ }\textbf {\bibinfo {volume} {95}},\ \bibinfo {pages} {014430} (\bibinfo {year} {2017}{\natexlab{a}})}\BibitemShut {NoStop}%
\bibitem [{\citenamefont {Labrune}\ and\ \citenamefont {Miltat}(1994)}]{topological}%
  \BibitemOpen
  \bibfield  {author} {\bibinfo {author} {\bibfnamefont {M.}~\bibnamefont {Labrune}}\ and\ \bibinfo {author} {\bibfnamefont {J.}~\bibnamefont {Miltat}},\ }\href@noop {} {\bibfield  {journal} {\bibinfo  {journal} {Journal of Applied Physics}\ }\textbf {\bibinfo {volume} {75}},\ \bibinfo {pages} {2156} (\bibinfo {year} {1994})}\BibitemShut {NoStop}%
\bibitem [{\citenamefont {Hierro-Rodriguez}\ \emph {et~al.}(2017{\natexlab{b}})\citenamefont {Hierro-Rodriguez}, \citenamefont {Quirós}, \citenamefont {Sorrentino}, \citenamefont {Valcárcel}, \citenamefont {Estébanez}, \citenamefont {Alvarez-Prado}, \citenamefont {Martín}, \citenamefont {Alameda}, \citenamefont {Pereiro}, \citenamefont {Vélez},\ and\ \citenamefont {Ferrer}}]{10.1063/1.4984898}%
  \BibitemOpen
  \bibfield  {author} {\bibinfo {author} {\bibfnamefont {A.}~\bibnamefont {Hierro-Rodriguez}}, \bibinfo {author} {\bibfnamefont {C.}~\bibnamefont {Quirós}}, \bibinfo {author} {\bibfnamefont {A.}~\bibnamefont {Sorrentino}}, \bibinfo {author} {\bibfnamefont {R.}~\bibnamefont {Valcárcel}}, \bibinfo {author} {\bibfnamefont {I.}~\bibnamefont {Estébanez}}, \bibinfo {author} {\bibfnamefont {L.~M.}\ \bibnamefont {Alvarez-Prado}}, \bibinfo {author} {\bibfnamefont {J.~I.}\ \bibnamefont {Martín}}, \bibinfo {author} {\bibfnamefont {J.~M.}\ \bibnamefont {Alameda}}, \bibinfo {author} {\bibfnamefont {E.}~\bibnamefont {Pereiro}}, \bibinfo {author} {\bibfnamefont {M.}~\bibnamefont {Vélez}},\ and\ \bibinfo {author} {\bibfnamefont {S.}~\bibnamefont {Ferrer}},\ }\href@noop {} {\bibfield  {journal} {\bibinfo  {journal} {Applied Physics Letters}\ }\textbf {\bibinfo {volume} {110}},\ \bibinfo {pages} {262402} (\bibinfo {year} {2017}{\natexlab{b}})}\BibitemShut {NoStop}%
\bibitem [{\citenamefont {Quiros}\ \emph {et~al.}(2018)\citenamefont {Quiros}, \citenamefont {Hierro-Rodriguez}, \citenamefont {Sorrentino}, \citenamefont {Valcarcel}, \citenamefont {Alvarez-Prado}, \citenamefont {Mart\'{\i}n}, \citenamefont {Alameda}, \citenamefont {Pereiro}, \citenamefont {V\'elez},\ and\ \citenamefont {Ferrer}}]{10261_173549}%
  \BibitemOpen
  \bibfield  {author} {\bibinfo {author} {\bibfnamefont {C.}~\bibnamefont {Quiros}}, \bibinfo {author} {\bibfnamefont {A.}~\bibnamefont {Hierro-Rodriguez}}, \bibinfo {author} {\bibfnamefont {A.}~\bibnamefont {Sorrentino}}, \bibinfo {author} {\bibfnamefont {R.}~\bibnamefont {Valcarcel}}, \bibinfo {author} {\bibfnamefont {L.~M.}\ \bibnamefont {Alvarez-Prado}}, \bibinfo {author} {\bibfnamefont {J.~I.}\ \bibnamefont {Mart\'{\i}n}}, \bibinfo {author} {\bibfnamefont {J.~M.}\ \bibnamefont {Alameda}}, \bibinfo {author} {\bibfnamefont {E.}~\bibnamefont {Pereiro}}, \bibinfo {author} {\bibfnamefont {M.}~\bibnamefont {V\'elez}},\ and\ \bibinfo {author} {\bibfnamefont {S.}~\bibnamefont {Ferrer}},\ }\href@noop {} {\bibfield  {journal} {\bibinfo  {journal} {Phys. Rev. Appl.}\ }\textbf {\bibinfo {volume} {10}},\ \bibinfo {pages} {014008} (\bibinfo {year} {2018})}\BibitemShut {NoStop}%
\bibitem [{\citenamefont {Sorrentino}\ \emph {et~al.}(2015)\citenamefont {Sorrentino}, \citenamefont {Nicol{\'{a}}s}, \citenamefont {Valc{\'{a}}rcel}, \citenamefont {Chich{\'{o}}n}, \citenamefont {Rosanes}, \citenamefont {Avila}, \citenamefont {Tkachuk}, \citenamefont {Irwin}, \citenamefont {Ferrer},\ and\ \citenamefont {Pereiro}}]{MISTRAL}%
  \BibitemOpen
  \bibfield  {author} {\bibinfo {author} {\bibfnamefont {A.}~\bibnamefont {Sorrentino}}, \bibinfo {author} {\bibfnamefont {J.}~\bibnamefont {Nicol{\'{a}}s}}, \bibinfo {author} {\bibfnamefont {R.}~\bibnamefont {Valc{\'{a}}rcel}}, \bibinfo {author} {\bibfnamefont {F.~J.}\ \bibnamefont {Chich{\'{o}}n}}, \bibinfo {author} {\bibfnamefont {M.}~\bibnamefont {Rosanes}}, \bibinfo {author} {\bibfnamefont {J.}~\bibnamefont {Avila}}, \bibinfo {author} {\bibfnamefont {A.}~\bibnamefont {Tkachuk}}, \bibinfo {author} {\bibfnamefont {J.}~\bibnamefont {Irwin}}, \bibinfo {author} {\bibfnamefont {S.}~\bibnamefont {Ferrer}},\ and\ \bibinfo {author} {\bibfnamefont {E.}~\bibnamefont {Pereiro}},\ }\href@noop {} {\bibfield  {journal} {\bibinfo  {journal} {Journal of Synchrotron Radiation}\ }\textbf {\bibinfo {volume} {22}},\ \bibinfo {pages} {1112} (\bibinfo {year} {2015})}\BibitemShut {NoStop}%
\bibitem [{\citenamefont {St{\"o}hr}(2023)}]{stöhr2023nature}%
  \BibitemOpen
  \bibfield  {author} {\bibinfo {author} {\bibfnamefont {J.}~\bibnamefont {St{\"o}hr}},\ }\href@noop {} {\emph {\bibinfo {title} {The Nature of X-Rays and Their Interactions with Matter}}},\ Springer Tracts in Modern Physics\ (\bibinfo  {publisher} {Springer International Publishing},\ \bibinfo {year} {2023})\BibitemShut {NoStop}%
\bibitem [{\citenamefont {Vansteenkiste}\ \emph {et~al.}(2014)\citenamefont {Vansteenkiste}, \citenamefont {Leliaert}, \citenamefont {Dvornik}, \citenamefont {Helsen}, \citenamefont {Garcia-Sanchez},\ and\ \citenamefont {Van~Waeyenberge}}]{Mumax}%
  \BibitemOpen
  \bibfield  {author} {\bibinfo {author} {\bibfnamefont {A.}~\bibnamefont {Vansteenkiste}}, \bibinfo {author} {\bibfnamefont {J.}~\bibnamefont {Leliaert}}, \bibinfo {author} {\bibfnamefont {M.}~\bibnamefont {Dvornik}}, \bibinfo {author} {\bibfnamefont {M.}~\bibnamefont {Helsen}}, \bibinfo {author} {\bibfnamefont {F.}~\bibnamefont {Garcia-Sanchez}},\ and\ \bibinfo {author} {\bibfnamefont {B.}~\bibnamefont {Van~Waeyenberge}},\ }\href@noop {} {\bibfield  {journal} {\bibinfo  {journal} {AIP Advances}\ }\textbf {\bibinfo {volume} {4}},\ \bibinfo {pages} {107133} (\bibinfo {year} {2014})}\BibitemShut {NoStop}%
\bibitem [{\citenamefont {Hierro-Rodriguez}\ \emph {et~al.}(2020)\citenamefont {Hierro-Rodriguez}, \citenamefont {Quirós}, \citenamefont {Sorrentino}, \citenamefont {Alvarez-Prado}, \citenamefont {Martín}, \citenamefont {Alameda}, \citenamefont {McVitie}, \citenamefont {Pereiro}, \citenamefont {Vélez},\ and\ \citenamefont {Ferrer}}]{NC2020}%
  \BibitemOpen
  \bibfield  {author} {\bibinfo {author} {\bibfnamefont {A.}~\bibnamefont {Hierro-Rodriguez}}, \bibinfo {author} {\bibfnamefont {C.}~\bibnamefont {Quirós}}, \bibinfo {author} {\bibfnamefont {A.}~\bibnamefont {Sorrentino}}, \bibinfo {author} {\bibfnamefont {L.}~\bibnamefont {Alvarez-Prado}}, \bibinfo {author} {\bibfnamefont {J.}~\bibnamefont {Martín}}, \bibinfo {author} {\bibfnamefont {J.}~\bibnamefont {Alameda}}, \bibinfo {author} {\bibfnamefont {S.}~\bibnamefont {McVitie}}, \bibinfo {author} {\bibfnamefont {E.}~\bibnamefont {Pereiro}}, \bibinfo {author} {\bibfnamefont {M.}~\bibnamefont {Vélez}},\ and\ \bibinfo {author} {\bibfnamefont {S.}~\bibnamefont {Ferrer}},\ }\href {https://doi.org/10.1038/s41467-020-20119-x} {\bibfield  {journal} {\bibinfo  {journal} {Nature Communications}\ }\textbf {\bibinfo {volume} {11}},\ \bibinfo {pages} {6382} (\bibinfo {year} {2020})}\BibitemShut {NoStop}%
\bibitem [{\citenamefont {Vogel}\ \emph {et~al.}(2011)\citenamefont {Vogel}, \citenamefont {Wintz}, \citenamefont {Gerhardt}, \citenamefont {Bocklage}, \citenamefont {Strache}, \citenamefont {Im}, \citenamefont {Fischer}, \citenamefont {Fassbender}, \citenamefont {McCord},\ and\ \citenamefont {Meier}}]{APLNIFE}%
  \BibitemOpen
  \bibfield  {author} {\bibinfo {author} {\bibfnamefont {A.}~\bibnamefont {Vogel}}, \bibinfo {author} {\bibfnamefont {S.}~\bibnamefont {Wintz}}, \bibinfo {author} {\bibfnamefont {T.}~\bibnamefont {Gerhardt}}, \bibinfo {author} {\bibfnamefont {L.}~\bibnamefont {Bocklage}}, \bibinfo {author} {\bibfnamefont {T.}~\bibnamefont {Strache}}, \bibinfo {author} {\bibfnamefont {M.-Y.}\ \bibnamefont {Im}}, \bibinfo {author} {\bibfnamefont {P.}~\bibnamefont {Fischer}}, \bibinfo {author} {\bibfnamefont {J.}~\bibnamefont {Fassbender}}, \bibinfo {author} {\bibfnamefont {J.}~\bibnamefont {McCord}},\ and\ \bibinfo {author} {\bibfnamefont {G.}~\bibnamefont {Meier}},\ }\href@noop {} {\bibfield  {journal} {\bibinfo  {journal} {Applied Physics Letters}\ }\textbf {\bibinfo {volume} {98}},\ \bibinfo {pages} {202501} (\bibinfo {year} {2011})}\BibitemShut {NoStop}%
\end{thebibliography}%

\end{document}